\documentclass[prl,
twocolumn,
superscriptaddress,showpacs,amssymb,amsmath,amsfonts,aps]{revtex4}

\usepackage{graphicx}
\usepackage{dcolumn}
\usepackage{epsfig}
\begin{document}

\title{Radiative decays of the $\Sigma^0(1385)$ and $\Lambda(1520)$ hyperons}

\newcommand{\bra}[1]{\left\langle #1 \right|}
\newcommand{\ket}[1]{\left| #1 \right\rangle}
\newcommand{\bracket}[2]{\left\langle #1 | #2 \right\rangle}
\newcommand{\matrixelement}[3]{\bra{#1} \hat{#2} \ket{#3}}
\newcommand\polvec{\vec\epsilon_\lambda^{\,*}(\vec k)}
\newcommand\etal{ \emph{et al.}\/}

\pacs{14.20.Jn,13.30.Ce,13.40.Hq}

\newcommand*{\ASU }{ Arizona State University, Tempe, Arizona 85287-1504} 
\affiliation{\ASU } 

\newcommand*{\SACLAY }{ CEA-Saclay, Service de Physique Nucl\'eaire, F91191 Gif-sur-Yvette,Cedex, France} 
\affiliation{\SACLAY } 

\newcommand*{\UCLA }{ University of California at Los Angeles, Los Angeles, California  90095-1547} 
\affiliation{\UCLA } 

\newcommand*{\CMU }{ Carnegie Mellon University, Pittsburgh, Pennsylvania 15213} 
\affiliation{\CMU } 

\newcommand*{\CUA }{ Catholic University of America, Washington, D.C. 20064} 
\affiliation{\CUA } 

\newcommand*{\CNU }{ Christopher Newport University, Newport News, Virginia 23606} 
\affiliation{\CNU } 

\newcommand*{\UCONN }{ University of Connecticut, Storrs, Connecticut 06269} 
\affiliation{\UCONN } 

\newcommand*{\DUKE }{ Duke University, Durham, North Carolina 27708-0305} 
\affiliation{\DUKE } 

\newcommand*{\ECOSSEE }{ Edinburgh University, Edinburgh EH9 3JZ, United Kingdom} 
\affiliation{\ECOSSEE } 

\newcommand*{\FIU }{ Florida International University, Miami, Florida 33199} 
\affiliation{\FIU } 

\newcommand*{\FSU }{ Florida State University, Tallahassee, Florida 32306} 
\affiliation{\FSU } 

\newcommand*{\GEISSEN }{ Physikalisches Institut der Universitaet Giessen, 35392 Giessen, Germany} 
\affiliation{\GEISSEN } 

\newcommand*{\GWU }{ The George Washington University, Washington, DC 20052} 
\affiliation{\GWU } 

\newcommand*{\ECOSSEG }{ University of Glasgow, Glasgow G12 8QQ, United Kingdom} 
\affiliation{\ECOSSEG } 

\newcommand*{\INFNFR }{ INFN, Laboratori Nazionali di Frascati, Frascati, Italy} 
\affiliation{\INFNFR } 

\newcommand*{\INFNGE }{ INFN, Sezione di Genova, 16146 Genova, Italy} 
\affiliation{\INFNGE } 

\newcommand*{\ORSAY }{ Institut de Physique Nucleaire ORSAY, Orsay, France} 
\affiliation{\ORSAY } 

\newcommand*{\ITEP }{ Institute of Theoretical and Experimental Physics, Moscow, 117259, Russia} 
\affiliation{\ITEP } 

\newcommand*{\JMU }{ James Madison University, Harrisonburg, Virginia 22807} 
\affiliation{\JMU } 

\newcommand*{\KYUNGPOOK }{ Kyungpook National University, Daegu 702-701, South Korea} 
\affiliation{\KYUNGPOOK } 

\newcommand*{\MIT }{ Massachusetts Institute of Technology, Cambridge, Massachusetts  02139-4307} 
\affiliation{\MIT } 

\newcommand*{\UMASS }{ University of Massachusetts, Amherst, Massachusetts  01003} 
\affiliation{\UMASS } 

\newcommand*{\MOSCOW }{ Moscow State University, General Nuclear Physics Institute, 119899 Moscow, Russia} 
\affiliation{\MOSCOW } 

\newcommand*{\UNH }{ University of New Hampshire, Durham, New Hampshire 03824-3568} 
\affiliation{\UNH } 

\newcommand*{\NSU }{ Norfolk State University, Norfolk, Virginia 23504} 
\affiliation{\NSU } 

\newcommand*{\OHIOU }{ Ohio University, Athens, Ohio  45701} 
\affiliation{\OHIOU } 

\newcommand*{\ODU }{ Old Dominion University, Norfolk, Virginia 23529} 
\affiliation{\ODU } 

\newcommand*{\PennState}{Penn State University, University Park, Pennsylvania 
16802
}
\affiliation{\PennState}

\newcommand*{\PITT }{ University of Pittsburgh, Pittsburgh, Pennsylvania 15260} 
\affiliation{\PITT } 

\newcommand*{\ROMA }{ Universita' di ROMA III, 00146 Roma, Italy} 
\affiliation{\ROMA } 

\newcommand*{\RPI }{ Rensselaer Polytechnic Institute, Troy, New York 12180-3590} 
\affiliation{\RPI } 

\newcommand*{\RICE }{ Rice University, Houston, Texas 77005-1892} 
\affiliation{\RICE } 

\newcommand*{\URICH }{ University of Richmond, Richmond, Virginia 23173} 
\affiliation{\URICH } 

\newcommand*{\SCAROLINA }{ University of South Carolina, Columbia, South Carolina 29208} 
\affiliation{\SCAROLINA } 

\newcommand*{\JLAB }{ Thomas Jefferson National Accelerator Facility, Newport News, Virginia 23606} 
\affiliation{\JLAB } 

\newcommand*{\UNIONC }{ Union College, Schenectady, NY 12308} 
\affiliation{\UNIONC } 

\newcommand*{\VT }{ Virginia Polytechnic Institute and State University, Blacksburg, Virginia   24061-0435} 
\affiliation{\VT } 

\newcommand*{\VIRGINIA }{ University of Virginia, Charlottesville, Virginia 22901} 
\affiliation{\VIRGINIA } 

\newcommand*{\WM }{ College of William and Mary, Williamsburg, Virginia 23187-8795} 
\affiliation{\WM } 

\newcommand*{\YEREVAN }{ Yerevan Physics Institute, 375036 Yerevan, Armenia} 
\affiliation{\YEREVAN } 

\newcommand*{\deceased }{ Deceased} 


\newcommand*{\NOWNCATU }{ North Carolina Agricultural and Technical State University, Greensboro, NC 27411}

\newcommand*{\NOWECOSSEG }{ University of Glasgow, Glasgow G12 8QQ, United Kingdom}

\newcommand*{\NOWSACLAY }{ CEA-Saclay, Service de Physique Nucl\'eaire, F91191 Gif-sur-Yvette,Cedex, France}

\newcommand*{\NOWSCAROLINA }{ University of South Carolina, Columbia, South Carolina 29208}

\newcommand*{\NOWJLAB }{ Thomas Jefferson National Accelerator Facility, Newport News, Virginia 23606}

\newcommand*{\NOWITEP }{ Institute of Theoretical and Experimental Physics, Moscow, 117259, Russia}

\newcommand*{\NOWOHIOU }{ Ohio University, Athens, Ohio  45701}

\newcommand*{\NOWFIU }{ Florida International University, Miami, Florida 33199}

\newcommand*{\NOWINFNFR }{ INFN, Laboratori Nazionali di Frascati, Frascati, Italy}

\newcommand*{\NOWCMU }{ Carnegie Mellon University, Pittsburgh, Pennsylvania 15213}

\newcommand*{\NOWINDSTRA }{ Systems Planning and Analysis, Alexandria, Virginia 22311}

\newcommand*{\NOWASU }{ Arizona State University, Tempe, Arizona 85287-1504}

\newcommand*{\NOWCISCO }{ Cisco, Washington, DC 20052}

\newcommand*{\NOWUK }{ University of Kentucky, LEXINGTON, KENTUCKY 40506}

\newcommand*{\NOWMOSCOW }{ Moscow State University, General Nuclear Physics Institute, 119899 Moscow, Russia}

\newcommand*{\NOWRPI }{ Rensselaer Polytechnic Institute, Troy, New York 12180-3590}

\newcommand*{\NOWDUKE }{ Duke University, Durham, North Carolina 27708-0305}

\newcommand*{\NOWUNCW }{ North Carolina}

\newcommand*{\NOWHAMPTON }{ Hampton University, Hampton, VA 23668}

\newcommand*{\NOWTulane }{ Tulane University, New Orleans, Lousiana  70118}

\newcommand*{\NOWORSAY }{ Institut de Physique Nucleaire ORSAY, Orsay, France}

\newcommand*{\NOWGEORGETOWN }{ Georgetown University, Washington, DC 20057}

\newcommand*{\NOWCUA }{ Catholic University of America, Washington, D.C. 20064}

\newcommand*{\NOWJMU }{ James Madison University, Harrisonburg, Virginia 22807}

\newcommand*{\NOWURICH }{ University of Richmond, Richmond, Virginia 23173}

\newcommand*{\NOWCALTECH }{ California Institute of Technology, Pasadena, California 91125}

\newcommand*{\NOWVIRGINIA }{ University of Virginia, Charlottesville, Virginia 22901}

\newcommand*{\NOWYEREVAN }{ Yerevan Physics Institute, 375036 Yerevan, Armenia}

\newcommand*{\NOWUMASS }{ University of Massachusetts, Amherst, Massachusetts  01003}

\newcommand*{\NOWRICE }{ Rice University, Houston, Texas 77005-1892}

\newcommand*{\NOWINFNGE }{ INFN, Sezione di Genova, 16146 Genova, Italy}

\newcommand*{\NOWROMA }{ Universita' di ROMA III, 00146 Roma, Italy}

\newcommand*{\NOWBATES }{ MIT-Bates Linear Accelerator Center, Middleton, MA 01949}

\newcommand*{\NOWVSU }{ Virginia State University, Petersburg,Virginia 23806}

\newcommand*{\NOWORST }{ Oregon State University, Corvallis, Oregon 97331-6507}

\newcommand*{\NOWGWU }{ The George Washington University, Washington, DC 20052}

\newcommand*{\NOWMIT }{ Massachusetts Institute of Technology, Cambridge, Massachusetts  02139-4307}

\author{S.~Taylor}

     \affiliation{\MIT}

\author{G.S.~Mutchler}

     \affiliation{\RICE}
  
\author{G.~Adams}

     \affiliation{\RPI}
\author{P.~Ambrozewicz}

     \affiliation{\FIU}
\author{E.~Anciant}

     \affiliation{\SACLAY}
\author{M.~Anghinolfi}

     \affiliation{\INFNGE}
\author{B.~Asavapibhop}

     \affiliation{\UMASS}
\author{G.~Asryan}

     \affiliation{\YEREVAN}
\author{G.~Audit}

     \affiliation{\SACLAY}
\author{H.~Avakian}

     \affiliation{\JLAB}
     \affiliation{\INFNFR}
% ...
 
\collaboration{The CLAS Collaboration}
\noaffiliation
\date{\today}

\begin{abstract}
The electromagnetic decays of the $\Sigma^0(1385)$ and $\Lambda(1520)$ hyperons
were studied in photon-induced reactions $\gamma p \rightarrow K^+ 
\Lambda(1116)\gamma$ in the CLAS detector 
at the Thomas Jefferson National Accelerator Facility. 
We report the first observation of the radiative decay of the 
$\Sigma^0(1385)$ and a measurement of the $\Lambda(1520)$ radiative decay width.
For the $\Sigma^0(1385)\rightarrow \Lambda(1116)\gamma$ transition, we measured a 
partial width of
$479\pm120(\mathrm{stat})^{+81}_{-100}(\mathrm{sys})$ keV, 
larger than all of the existing model predictions.  For the 
$\Lambda(1520)\rightarrow\Lambda(1116)\gamma$ transition, we obtained a partial 
width of $167\pm43(\mathrm{stat})^{+26}_{-12}(\mathrm{sys})$ keV.
\end{abstract}

\maketitle

\section{Introduction}
The low-lying neutral excited-state hyperons $\Sigma^0(1385)$, 
$\Lambda(1405)$, and $\Lambda(1520)$ were discovered in the 1960s, but their 
quark wave functions are still not well-understood and experimental studies 
of their properties have been scarce since the early 1980s. 
The electromagnetic decays of baryons produced in photon reactions provide an especially
clean method of probing their wave functions. Baryons with a strange quark have an 
additional degree of freedom which aids in the study of multiplet mixing and non 
3-quark admixtures. Recently there has been a renewal of interest in this field,
e.g. electro-production of the $\Lambda(1520)$\cite{Barrow:2001ds}.  
This paper reports the results of a non-model dependent 
measurement of the radiative decay of the $\Sigma^0(1385)$ and $\Lambda(1520)$.

\begin{figure}
\epsfig{file=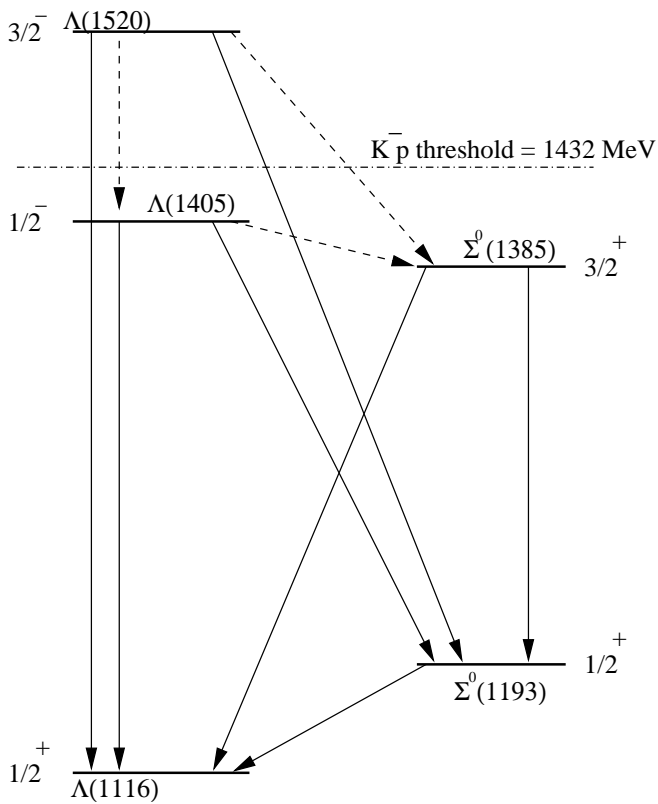,width=\columnwidth}
\caption{Photon decay spectrum of low lying excited state hyperons. The transitions shown as dashed lines are suppressed.  
}
\label{hierarch}
\end{figure}
The non-relativistic quark model (NRQM) of Isgur and 
Karl\cite{isgur} has been remarkably successful in predicting the masses and
widths of $N^*$ and $\Delta^*$ states, but less successful in the strange 
sector.  
Several competing
models for hyperon wave functions have been proposed.   
Measuring 
the transitions $Y \to \Lambda(1116) \gamma$ and $Y \to \Sigma(1193) \gamma$
provides a means of differentiating between these models.
Calculations have
been done in the framework of NRQM\cite{kaxiras,DHK}, a relativized 
constituent quark model 
(RCQM)\cite{warns}, a chiral constituent quark model
($\chi$CQM) that includes 
electromagnetic exchange currents between the quarks\cite{wagner},
 the MIT bag model\cite{kaxiras}, the chiral bag model\cite{umino}, 
the bound-state soliton model\cite{Schat}, a three-flavor generalization of 
the Skyrme model that uses the collective approach instead of the bound-state 
approach\cite{Abada:1995db,Haberichter:1996cp},
an algebraic model of hadron structure\cite{Bijker:2000gq},  
heavy baryon chiral perturbation 
theory (HB$\chi$PT)\cite{butler}, and the $1/N_c$ expansion of 
QCD\cite{Lebed:2004zc}.
The radiative widths in keV are tabulated in Table \ref{widths}. 
The $\Delta\rightarrow p\gamma$ width is 
included for comparison.  

The photon decay spectrum of the low-lying excited state hyperons is shown in 
Fig. \ref{hierarch}.
The widths given in Table \ref{widths} can be qualitatively estimated using SU(3) symmetry. The 
$\Lambda(1116)$ and the $\Sigma^0(1193)$ are in the S=$1/2^+$ SU(3) octet and the 
$\Sigma^{0}(1385)$ is in the S=$3/2^+$ SU(3) decuplet. 
The $\Lambda(1116)$ has the two light quarks in the s orbital in a spin S=0, isospin T=0 configuration. The
$\Sigma^0(1193)$ and the $\Sigma^0(1385)$ have the light quarks in a spin S=1, T=1 configuration. 
All three hyperons have the strange quark in the s orbital.
Decuplet to octet radiative decays are dominated by
an M1 transition with a spin-flip of one quark.  
The SU(3) model prediction of the ratio $\Sigma^* \rightarrow \Sigma \gamma$ to the $\Sigma^*
\rightarrow \Lambda \gamma $ is $\sim\frac{1}{6}$ times kinematic factors.   
This, plus the fact that most of the 
constituent quark model calculations\cite{kaxiras,DHK,Koniuk:vy,warns,wagner} listed 
in Table \ref{widths} used the impulse approximation, leads to a very narrow range of 
predictions, (265--273 keV) for the $\Sigma^*
\rightarrow \Lambda \gamma$ reaction and    
(17.4--23 keV) for the $\Sigma^* \rightarrow \Sigma \gamma$ reaction.
The $\Lambda(1405)$ and $\Lambda(1520)$ have light quarks in the s orbital with S=0, T=0
and the strange quark in a p$\frac{1}{2}$ and p$\frac{3}{2}$ orbital respectively. 
The  radiative decays $\Lambda(1520) \rightarrow \Sigma^*, \Sigma$ and 
$\Lambda(1405) \rightarrow \Sigma^*, \Sigma$ require that the strange quark make a transition 
from a p orbital to an s orbital with a simultaneous spin flip of one of the light quarks. 
These transitions 
are thus forbidden by the one body nature of the electromagnetic operator. They can proceed only 
via configuration mixing introduced by, \emph{e.g.}\/ the QCD hyperfine interaction, which leads to a 
wider range of model predictions. This is explained 
in more detail in an excellent review of the experimental and theoretical situation in \cite{Lands}. 
\begin{table*}
\begin{center}
\begin{tabular}{l|c|cc|cc|cc}
&$\Delta(1232)$&\multicolumn{2}{c|}{$\Sigma^0(1385)$} & \multicolumn{2}{c|}{$\Lambda(1405)$} & 
\multicolumn{2}{c}{$\Lambda(1520)$}\\  
Model & $p\gamma$ & $\Lambda(1116)\gamma$ & $\Sigma^0(1193)\gamma$ & $\Lambda(1116)\gamma$ & $\Sigma^0(1193)\gamma$ &
$\Lambda(1116)\gamma$ & $\Sigma^0(1193)\gamma$ \\\hline\hline
NRQM\cite{kaxiras,DHK} & 360\cite{Koniuk:vy}  & 273	& 22	& 200	& 72	& 156	& 55\\
RCQM\cite{warns}   &  & 267	& 23	& 118	& 46	& 215	& 293\\
$\chi$CQM\cite{wagner} & 350  & 265 & 17.4 &  & & & \\
\hline
MIT Bag\cite{kaxiras}  &  & 152	& 15	& 60, 17	& 18, 2.7& 46	& 17\\
Chiral Bag\cite{umino}& &  	&	& 75	& 1.9	& 32 	& 51\\
\hline
Soliton\cite{Schat} &  	& 243, 170 & 19, 11 & 44, 40 & 13, 17 &&\\ 
Skyrme\cite{Abada:1995db,Haberichter:1996cp}& 309-348  & 157-209 & 7.7-16 & & & & \\
\hline
Algebraic model\cite{Bijker:2000gq} & 343.7 & 221.3 & 33.9 & 116.9 & 155.7 & 85.1 & 180.4 \\
\hline
HB$\chi$PT\cite{butler}$^\dag$ & (670-790)  & 290-470 & 1.4-36&	&	&	&\\\hline
$1/N_c$ expansion\cite{Lebed:2004zc} &  & $298\pm25$ & $24.9\pm4.1$ & & & \\
\hline\hline
Previous Experiments 	 & 640-720\cite{pdg}  
&$<$2000\cite{colas}    & $<$1750\cite{colas}  &  27$\pm$8\cite{Burkhardt} 
	& 10$\pm$4\cite{Burkhardt} 
&  33$\pm$11\cite{bertini}
	& 47$\pm$17\cite{bertini} \\
&    & & & &  23$\pm$7\cite{Burkhardt}  & 134$\pm$23\cite{mast} & \\
&    & & & &                            & 159$\pm$33$\pm$26\cite{Antipov:2004qp}&
\\\hline
This experiment  &     &  $479\pm120^{+81}_{-100}$ & & & & 
$167\pm43^{+26}_{-12}$ & 
\end{tabular}   
\end{center}
\caption{Theoretical predictions and experimental values for the radiative 
widths (in keV) for the transitions $Y\rightarrow\Lambda(1116)\gamma$
and $Y\rightarrow\Sigma(1193)\gamma$. Some models have multiple predictions
that depend on different assumptions.   For comparison the predictions and 
experimental value are quoted for the $\Delta(1232)\rightarrow p \gamma$ 
transition. $^\dag$The results for HB$\chi$PT\cite{butler} are normalized to 
the quoted empirical range (in parentheses) for the 
$\Delta\rightarrow p \gamma$ transition. 
}
\label{widths}
\end{table*}

Experimental measurements have 
been sparse.  The results are tabulated in Table \ref{widths}.
The $\Lambda(1520)\rightarrow\Lambda\gamma$ transition has been measured by 
Mast\etal\cite{mast} 
using a $K^-$ beam  with a liquid-hydrogen bubble chamber, by
 Bertini\etal\cite{bertini} (unpublished) with a liquid-hydrogen target viewed by 
a  NaI detector, and by Antipov\etal\cite{Antipov:2004qp} using a
 high-energy proton beam on carbon and copper targets.  Antipov\etal\ measured the $K^+$, p and $\pi^-$ in a magnetic spectrometer and  
detected the decay photons using an electromagnetic calorimeter. 
These are the only direct measurements in the literature. 
Burkhardt and Lowe\cite{Burkhardt} extracted model dependent branching ratios for 
$\Lambda(1405)$ radiative decay from the kaon-proton capture data 
of Whitehouse\etal\cite{whitehouse}.  
The radiative decay of the $\Sigma^0(1385)$ has never been observed 
(Meisner\cite{meisner} reports one event); only 
upper limits for the branching ratios have 
been established\cite{colas}. 

\section{Experiment}

In the current experiment, the low-lying excited-state hyperons were 
studied in the reaction $\gamma p \rightarrow K^+ p \pi^-X$ using the 
CEBAF Large Acceptance
Spectrometer (CLAS) in Hall B at the Thomas Jefferson National Accelerator
 Facility. The data were from the G1C running period September to October 1999. 
The primary electron beam was converted to a 
photon beam with a thin radiator of $10^{-4}$ radiation lengths. The 
scattered electron is momentum-analyzed by a photon tagging 
spectrometer\cite{tagnim} with a resolution of $\Delta$E/E = $10^{-3}$.  
Photons were tagged over a range of 20--95\% of the incident electron beam 
energy. The electron beam energies were 
2.445 GeV, 2.897 GeV, and  3.115 GeV and the currents were typically 6 nA.
The target was liquid hydrogen in a cylindrical cell of 17.9 cm 
length and 2 cm radius.  The 
CLAS detector\cite{mecking} consists of six individually 
instrumented segments, each consisting of three layers of drift chambers
and a shell of 48 time-of-flight scintillators. Six superconducting magnets
provided a toroidal magnetic field, with negative particles 
bent toward the beam direction. 
The trigger consisted of a triple 
coincidence between the photon tagger, the time-of-flight system, and a small 
scintillation detector (the ``Start Counter''\cite{startnim}) surrounding 
the target scattering chamber. Only one charged particle in the CLAS was required in
the trigger to accommodate the 6 experiments that were running simultaneously.
A total of 1420M triggers were collected at 2.445 GeV, 845M at 2.897 and 2280M at 3.115 GeV.

\subsection{Particle Identification}

\begin{figure}
\begin{center}
\epsfig{file=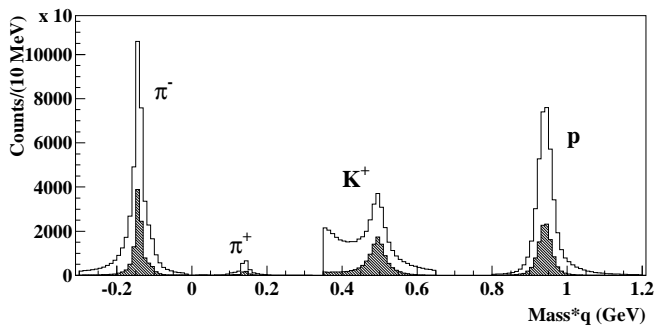,width=\linewidth}
\end{center}
\caption{ Particle Identification:  Hadron mass from TOF and momentum 
information multiplied by the sign of the charge of the particle. The shaded curve
is the mass spectrum after the PID cuts.  
}
\label{pid}
\end{figure}

Charged hadrons were identified using momentum and time-of-flight information.
The processed data files were filtered for events containing one $K^+$, one $\pi^-$, and one proton track
in coincidence with the incident tagged photon.
Kaon candidates were chosen using a broad range in mass (0.35--0.65 GeV).  
The $\pi^-$ candidates were selected with a mass range of
 $<$0.3 GeV and proton candidates with a range of 0.8--1.2 GeV.
A minimum momentum cut of 0.3 GeV/c was applied for the kaons and protons and   
0.1 GeV/c for the pions.    
 The hadron mass spectrum for events that survive the filter is shown in 
Fig. \ref{pid}.
The kaon peak sits on top of a large background due to high momentum pions with poorly determined mass.

To further refine the kaon identification the 
difference $\Delta t$ between the time at the target for the kaon candidate
and what it would be for a true kaon was computed:
\begin{equation}
\Delta t = (t_{TOF} - t_{vert}) \left(1-\sqrt{\frac{p^2+M_{K^+}^2} 
	{p^2+M_{calc}^2}} \right),
\end{equation}
\begin{equation}
 \ M_{calc}^2= \frac{p^2}{\gamma^2 \beta^2}, 
\end{equation}
where $t_{TOF} - t_{vert}$ is the flight time between the nteraction vertex and 
the he Time-of-Flight array. 
We require $|\Delta t | < 0.67$ ns.
Since the experiment consists of two physically separate systems, the Tagger and the CLAS detector, 
we require that the time at the interaction vertex measured by the two systems 
agree to within $5\%$ of the flight time between 
the Start Counter and the TOF paddles.

The CLAS detector does not cover the full angular range in $\theta$ or $\phi$.
Some angular regions are shadowed by the toroidal coils.   The shadow region 
broadens in $\phi$ as a function of decreasing $\theta$ as seen from the center of the 
target.  
  All tracks were required to be in the region of well-understood acceptance 
by applying a fiducial cut of the form
\begin{equation}
\theta>4.0 + \frac{510.58}{(30-\phi)^{1.5518}},
\end{equation}
where $\phi$ is the azimuthal angle folded onto the range 0--30$^\circ$.
We also require $\phi<26^\circ$.

\begin{figure}
\begin{center}
\epsfig{file=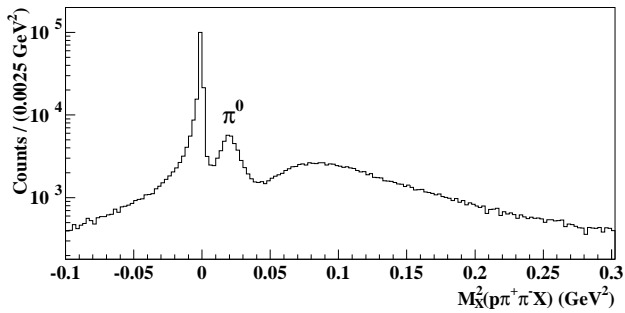,width=\linewidth}
\end{center}
\caption{Pion contamination:  Mass squared ($M_X^2$) 
for the $\gamma p \rightarrow p \pi^+ \pi^- (X) $
reaction where the $\pi^+$ was a potentially mis-identified kaon.}
\label{pion_contam}
\end{figure}

Some of the "kaon" events are really misidentified $\pi^+$. This can be seen in Fig. \ref{pion_contam}\
 where all events are plotted assuming 
that all kaon candidates are really  misidentified $\pi^+$ 
and compute the missing mass squared for the reaction 
$\gamma p \rightarrow p\pi^+\pi^-(X)$.  The prominent 
spike at zero mass squared indicates $\gamma p \rightarrow p\pi^+\pi^-$ 
contamination and a $\pi^0$ peak is clearly evident but at a much reduced 
level.  The expected distribution for good $K^+$ events goes 
to zero for zero $p\pi^+\pi^-$ missing mass squared. We require the missing 
mass squared from this calculation to be 
greater than 0.01 $GeV^2$ to eliminate for example $\rho\rightarrow\pi^+\pi^-$
contamination.  We did not cut above the 
$\pi^0$ peak in Fig. \ref{pion_contam} because that would have cut into the good 
$K^+$ events.
The hadron mass spectrum after all of the above cuts have been applied 
are shown as the shadowed histogram in 
Fig. \ref{pid}.

 The kaon momentum is corrected for average dE/dx losses in the target 
material, target wall, 
carbon epoxy pipe and the start counter depending on the position of the 
primary vertex, which is approximated by the intersection of the proton and kaon tracks.
The ground-state $\Lambda$ is sufficiently long-lived that it decays a 
measurable distance from the primary vertex.  The secondary vertex is 
determined by the intersection of the proton and $\pi^-$ tracks.  The proton 
and $\pi^-$ tracks are corrected for average dE/dx losses according to the position of the 
secondary vertex.
 
\begin{figure}[t]
\begin{center}
\epsfig{file=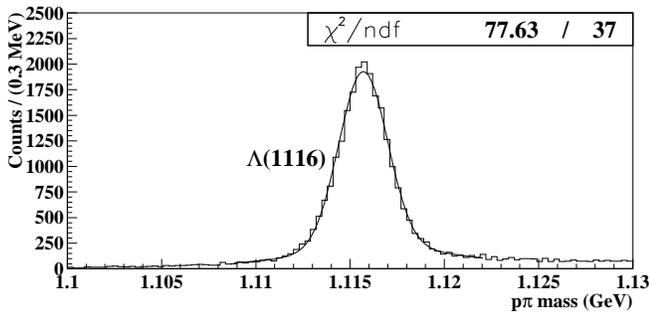,width=\linewidth}
\end{center}
\caption{$\Lambda$ identification: proton-$\pi^-$ invariant mass}
\label{lambda}
\end{figure}

\begin{figure}[b]
\begin{center}
\epsfig{file=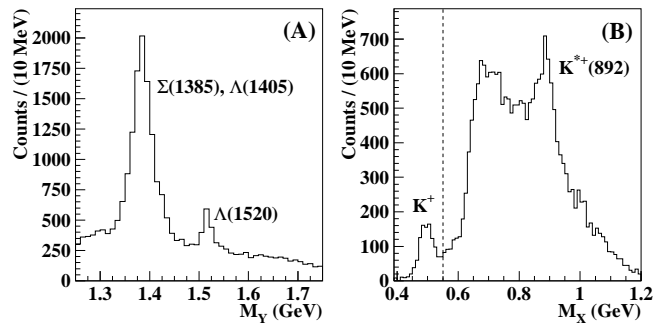,width=\linewidth}
\end{center}
\caption{(A) Missing mass for the reaction $\gamma p \rightarrow K^+(X)$.
 (B) Missing mass for the reaction $\gamma p \rightarrow \Lambda (X)$.
}
\label{ystar}
\end{figure}

\begin{figure}[b]
\begin{center}
\epsfig{file=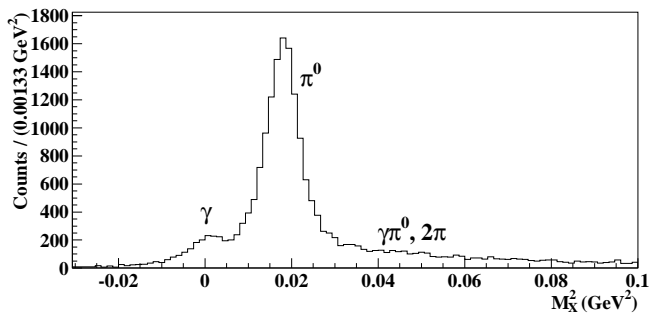,width=\linewidth}
\end{center}
\caption{Missing mass squared for the reactions $\gamma p \rightarrow K^+ 
\Lambda (X)$.}
\label{missing}
\end{figure}

The four-momentum of the $\Lambda(1116)$ was reconstructed from the
proton and $\pi^-$ four-momenta (Fig. \ref{lambda}).  The Gaussian  
resolution of the $\Lambda$ peak is about $\sigma=1.3$ MeV,  
consistent with the instrumental resolution.
The excited-state hyperon mass spectrum  for the region between 1.25 GeV and 1.75 GeV 
requiring the $p\pi^-$ invariant mass to be in the 
range 1.112--1.119 GeV is shown 
in Fig. \ref{ystar}A.  Fig. \ref{ystar}B shows the mass $M_X$ from the
reaction $\gamma p \rightarrow \Lambda (X)$.  A clear peak at the mass of the 
$K^*(892)$ is seen. 
The peak at the $K^+$ mass is due to accidentals under the TOF peak.
This background is eliminated by requiring $M_X>0.55$ GeV.
Fig. \ref{missing} shows the missing mass squared for the
reaction $\gamma p \rightarrow K^+ \Lambda (X)$ after the foregoing cuts have 
been applied.   A prominent peak shows up at
$M_{\pi^0}^2$ and a smaller peak at zero missing mass squared. 
The counts above the $\pi^0$ peak  are typically due to 
$\gamma p \rightarrow K^+\Sigma^0(X)$.

\subsection{Kinematic fitting}
\label{kinematic_fitting}

A better approximation to the primary and secondary vertices can be found using kinematic fitting. 
We used the Lagrange multiplier method \cite{frodesen}. 
The unknowns are divided into a set of measured variables ($\vec\eta$) 
and a set of unmeasured variables ($\vec\xi$) 
such as the missing momentum or the 4-vector for a decay particle.  
For each constraint equation a Lagrange multiplier $\lambda_i$ is introduced.  
We minimize
\begin{equation}
\chi^2(\vec\eta,\vec\xi,\vec\lambda)= (\vec\eta_0-\vec\eta)^T V^{-1} 
	(\vec\eta_0-\vec\eta)+ 2 \vec\lambda^T \vec f 
\end{equation}
by differentiating $\chi^2$ with respect to all the variables, linearizing the 
constraint equations and iterating.  Here $\vec\eta_0$ is a vector containing 
the initial guesses for the measured quantities and $V$ is the covariance 
matrix comprising the estimated errors on the measured quantities.
We iterate until the difference in
magnitude between the current $\chi^2$ and the previous value is $\leq$0.001. 
The covariance matrix $V$ for each track returned by the tracking code does 
not contain the effects of multiple scattering and energy loss in the target 
cell, 
the carbon epoxy pipe, or the start counter. To correct for this we apply
multiple scattering and energy loss corrections 
to the diagonal matrix elements.  

\begin{figure}
\begin{center}
\epsfig{file=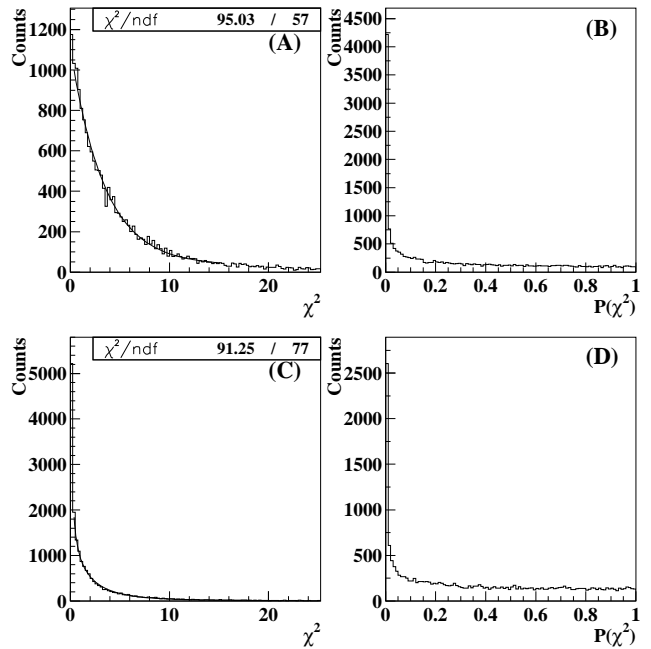,width=\linewidth}
\end{center}
\caption{$\chi^2$ and Confidence Level distributions for the $\Lambda\rightarrow p \pi^-$ fit 
(A and B) and 
the $K^+\Lambda$ vertex fit (C and D).}
\label{chi2lam}
\end{figure}

The first step in the fitting procedure is to fit the proton and $\pi^-$ 
tracks with the $\Lambda$
hypothesis.  This is a 2C fit.  There are six unknowns ($\vec p_\Lambda$,
$\vec r_{V2}$) and eight constraint equations, 
\begin{equation}
\vec f = \left[ \begin{array}{c}
	E_p+E_\pi-E_\Lambda \\
	\vec p _p + \vec p _\pi - \vec p_\Lambda \\
	(y-y_\pi) p^z_\pi-(z-z_\pi)p^y_\pi\\
        (x-x_\pi) p^z_\pi-(z-z_\pi)p^x_\pi\\
        (y-y_p) p^z_p-(z-z_p)p^y_p\\
        (x-x_p) p^z_p-(z-z_p)p^x_p\\
	\end{array} \right]=\vec 0. 
\end{equation} 
The $\chi^2$ distribution for this fit is shown in Fig. \ref{chi2lam}A
and the Confidence Level plot is shown in Fig. \ref{chi2lam}B.
The curve is the result of a fit to the histogram using the function form 
of a $\chi^2$ distribution with two degrees of freedom plus a flat background 
term. Explicitly,
\begin{equation}
f(\chi^2)= \frac{P_1}{2} e^{-P_2 \chi^2/2}+P_3.
\end{equation}
 The fit result ($P_2=0.558$)
 suggests that we are underestimating the errors in the proton and 
$\pi^-$ tracks, but the shape is close to the expected shape.
The Confidence Level is given by the equation
\begin{equation}
CL = \int_{\chi^2}^{\infty}f(z;n)dz
\end{equation}
where f(z;n) is the $\chi^2$ probability density function with n degrees of freedom.

The second step is to use these Kaon and Lambda tracks 
to obtain a better primary vertex. 
This is a 1C fit.  There are 3 unknowns ($\vec r_{V1}$) and four
constraint equations.  The  $\chi^2$ distribution for this fit is shown in 
Fig. \ref{chi2lam}C and the Confidence Level plot is shown in Fig. \ref{chi2lam}D.
 The curve in Fig. \ref{chi2lam}C is the result of a 
fit to the histogram  using the functional form of a $\chi^2$ distribution 
with one degree of freedom plus a flat background term.  Explicitly,
\begin{equation}
f(\chi^2)=\frac{P_1}{\sqrt{2}\ \Gamma(1/2)}\frac{e^{-P_2 \chi^2/2}} 
{ \sqrt{\chi^2}} + P_3.
\label{chi2_1}
\end{equation}
with a fit result of $P_2=0.507$.
We require the probability of 
the $\Lambda\rightarrow p \pi^-$ fit and the primary vertex fit be    
 $\le0.5$\% of exceeding $\chi^2$ for an ideal $\chi^2$ distribution.
The improved kaon and lambda four vectors are used to compute the excited-state 
hyperon mass spectrum and the missing mass squared.  

\begin{figure}
\begin{center}
\epsfig{file=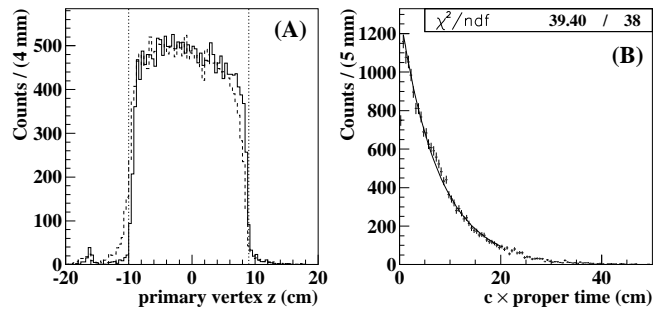,width=\linewidth}
\end{center}
\caption{ A. Z-position of primary vertex.  Solid histogram:  K$\Lambda$ fit.  
Dashed histogram:  Kp fit. 
B. Lambda decay proper time in units of $cm$. The excited-state hyperon 
mass was greater than 1.25 GeV for both plots.}
\label{primary_vertex}
\end{figure}

\begin{figure}
\begin{center}
\epsfig{file=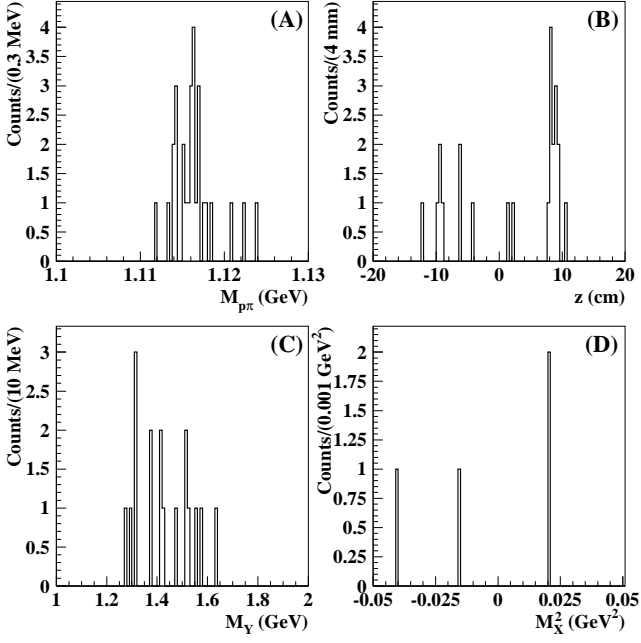,width=\linewidth}
\end{center}
\caption{Empty target results. A. Proton piminus invariant mass. B. Vertex z-position.
C. Hyperon mass. D. Missing mass squared for $M_Y$=1.34--1.43 GeV.}
\label{empty}
\end{figure}

Fig. \ref{primary_vertex}A  
compares the z-position of the primary vertex from the improved fitting procedure to the naive
 kaon-proton result.
We apply a target z-position 
cut for the primary vertex between 
-10.0 cm and +9.0 cm and a radial cut of 2 cm.  These cuts were chosen to 
ensure that the primary event came from the target region.
The proper time of the $\Lambda$ decay is plotted in Fig. \ref{primary_vertex}B. 
An exponential fit to the data gives a decay constant of $7.62\pm 0.09$ cm which is
comparable to the PDG value of 7.89$\pm$0.06 cm.
To verify that the target walls do not make a 
significant contribution to our yields, we applied the analysis procedure 
described above to the empty-target data. 
For the empty target runs the beam current ranged between 10 and 24 nA and
averaged about 15 nA.
The results from analyzing about 33 million empty-target events (corresponding to approximately 
$\frac{1}{3}$ of the target full integrated photon flux) are shown in 
Fig. \ref{empty}.  We obtained 25 $\Lambda(1116)$ candidates within the
proton-$\pi^-$ invariant mass range of 1.112--1.119 GeV (Fig. \ref{empty}A). 
The z distribution is shown in Fig. \ref{empty}B. 
The hyperon mass distribution for those events satisfying the
vertex cut is shown 
in Fig. \ref{empty}C.  Figure \ref{empty}D shows the missing mass squared 
distribution for hyperon masses in the 1.34--1.43 GeV range. There are no counts
near zero missing mass squared and only two near $m_{\pi^0}^2$. Both of these
counts have z and r positions within the target volume.
They correspond to interactions with the residual (cold) hydrogen gas in the 
target.  From this we conclude that the background due to interactions with 
the walls of the target cell is negligible.

To achieve $\gamma/\pi^0$ separation, the events were sorted   
according to topology using 
kinematic fits with two hypotheses
\begin{center}
\begin{tabular}{ccc}
R1:&$\gamma p \rightarrow K^+ \Lambda \pi^0$  & 1C\\
R2:&$\gamma p \rightarrow K^+ \Lambda \gamma$ & 1C \\
\end{tabular}
\end{center}
The corresponding constraint equations are
\begin{equation}
\vec f = \left[ \begin{array}{c}
	E_{beam}+M_p-E_K-E_\Lambda-E_X\\
	\vec p_{beam} - \vec p _K - \vec p_\Lambda -\vec p_X
	\end{array} \right]=\vec 0. 
\end{equation}
Here X is a missing $\pi^0$ or  a missing $\gamma$.

\begin{figure}
\begin{center}
\epsfig{file=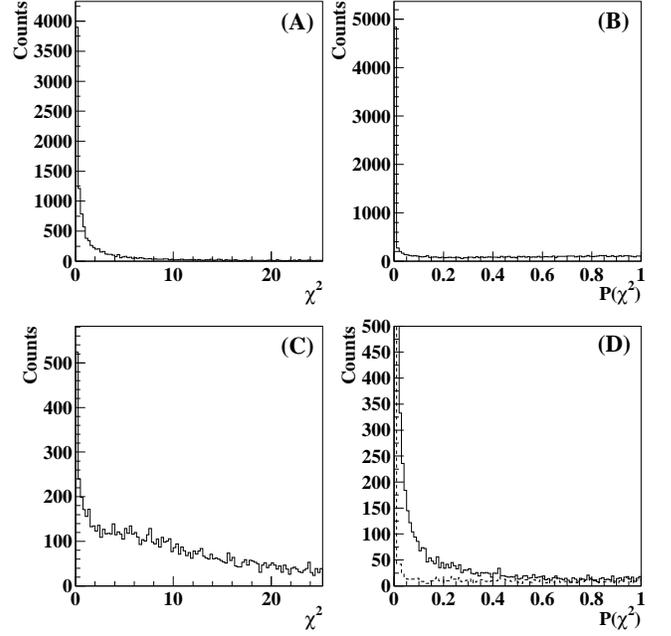,width=\linewidth}
\end{center}
\caption{$\chi^2$ and Confidence Level distributions for the two reactions R1 (A and B) and R2 (C and D).
The dashed curve in D is the R2 Confidence Level with the R1 reaction vetoed with $\chi^2$=3.841.}
\label{chi2}
\end{figure}

The $\chi^2$ distributions for reactions R1 and R2 are shown in Fig. 
\ref{chi2}A and \ref{chi2}C, respectively.
 The hyperon mass range was 1.25--1.75~GeV.   The corresponding Confidence Levels plots 
are shown in 
Fig. \ref{chi2}B and \ref{chi2}D.
For R1 we obtain the expected shape for a $\chi^2$ distribution with one degree of 
freedom.  For R2 the 
$\chi^2$ values indicates that the radiative decay hypothesis is 
inconsistent with most of the events. The dashed curve in Figure \ref{chi2}D is
the Confidence Level for hypothesis R2 for those events which do not satisfy hypothesis R1 
at the 5\% level. 
We now see a shape consistent with a $\chi^2$ distribution with one 
degree of freedom.   

\begin{figure}
\begin{center}
\epsfig{file=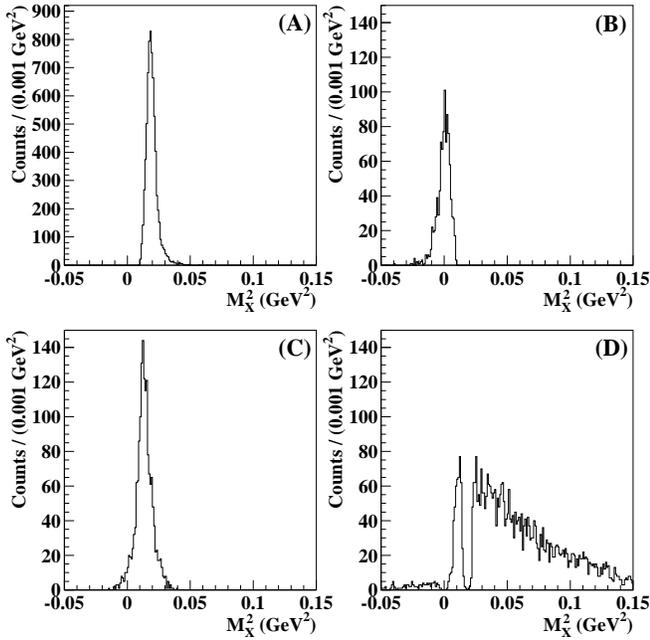,width=\linewidth}
\end{center}
\caption{Missing mass squared distributions for $\chi^2_{HIGH} = \chi^2_{LOW}$ =3.841.   
Cases A--D are explained in the text.
}
\label{missing_cuts}
\end{figure}

\begin{figure}
\begin{center}
\epsfig{file=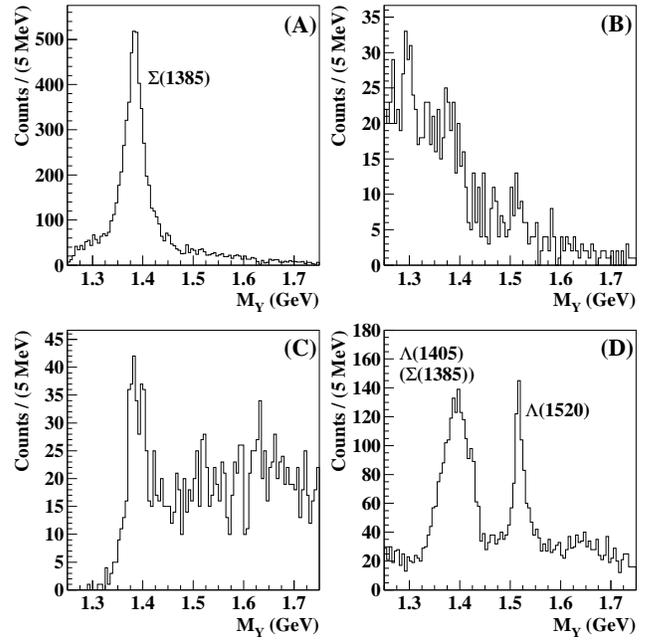,width=\linewidth}
\end{center}
\caption{Hyperon mass distributions for $\chi^2_{HIGH} = \chi^2_{LOW}$ = 3.841. 
Cases A-D are explained in the text.
}
\label{ystar_cuts}
\end{figure}

Fig. \ref{missing_cuts} shows the missing mass squared distributions for
a representative set of $\chi^2$ cuts.  For the purpose of the plot, we 
require 
$\chi^2_{R1}\geq3.841$ and $\chi^2_{R2}<3.841$ to isolate the radiative 
channel (case B).  To isolate
the pion channel (case A), we require $\chi^2_{R1}<3.841$ and  
$\chi^2_{R2}\geq3.841$.  Case C is the ``ambiguous'' case where both 
$\chi^2_{R1}<3.841$ and $\chi^2_{R2}<3.841$. 
 Case D consists of those events 
that do not agree with either the radiative channel or the pion channel,
for which $\chi^2_{R1}\ge 3.841$ and $\chi^2_{R2}\ge 3.841$.
For a 1C fit $\chi^2=3.841$ corresponds to a 5\% probability of exceeding 
$\chi^2$ for an ideal $\chi^2$ distribution.
The ``ambiguous'' events are most likely to  
be $\gamma p \rightarrow K^+\Lambda\pi^0$ events.  
Case D events are most likely 
be $\gamma p \rightarrow K^+\Sigma^0\pi^0$ events.  
Fig. \ref{ystar_cuts} shows the 
corresponding hyperon mass spectra. 
Fig. \ref{ystar_cuts}A is dominated
by the $\Sigma^0(1385)\rightarrow\Lambda\pi^0$ channel, for which the branching 
ratio is $\sim$88\%\cite{pdg}.  
We calculated the $\Sigma(1385)$ radiative transition relative to 
this channel.
The $\Lambda(1520)$ peak shows up in 
Fig. \ref{ystar_cuts}D because of the decay channels 
$\Lambda(1520)\rightarrow \Sigma^0\pi^0$ (BR=14\%) and 
$\Lambda(1520)\rightarrow\Lambda\pi\pi$ (BR=10\%).

\subsection{Double Bremsstrahlung}
\label{db}

\begin{figure}
\begin{center}
\epsfig{file=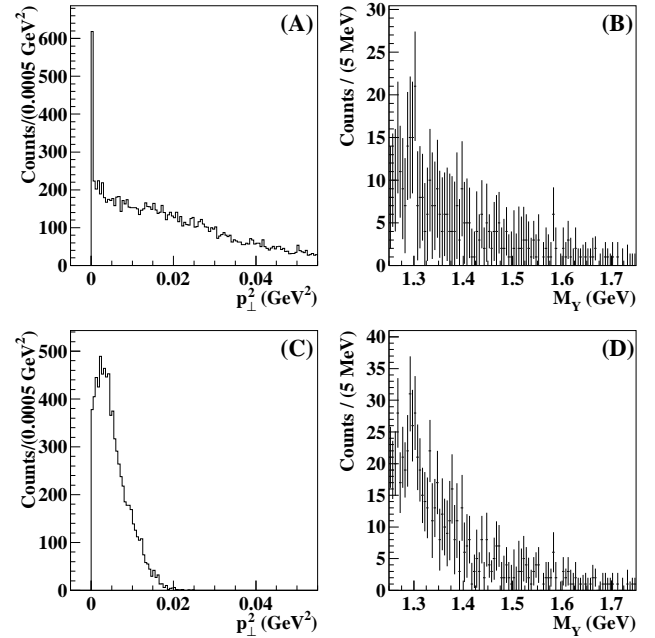,width=\linewidth}
\end{center}
\caption{Effect of $p_\perp^2$ cut on $\gamma$ channel.
A. The $p_\perp^2$ momentum spectrum for $K^+ \Lambda(X)$ events.
 B. The $\gamma$ channel cut distibution for a 0.0004 GeV$^2$ cut.
C. The $p_\perp^2$ momemtum spectrum for $K^+ \Sigma^0(X)$ events.
D.  The $\gamma$ channel cut distibution for a 0.015 GeV$^2$ cut. }
\label{pperp}
\end{figure}

The $\gamma$ channel does not show the structure expected from hyperon photon
decays. 
The structure was found to be masked by a background resulting from double bremsstahlung in the radiator.
The reaction 
$\gamma_1 + \gamma_2 p \rightarrow K^+ \Lambda + \gamma_1$ can mimic the reaction
$\gamma p \rightarrow K^+ \Lambda \gamma$. 
But in this case the missing momentum 
from the reaction $\gamma p \rightarrow K^+ \Lambda(X)$ points along the 
+z direction (along the beam).  
This can also happen if the event is an accidental or 
inefficiencies in the tagger plane allow the wrong electron to be selected. 
This problem is illustrated in Fig. 
\ref{pperp}.  Fig. \ref{pperp}A shows the off-z-axis 
momentum $p_\perp^2=p_x^2+p_y^2$ for the candidate missing particle.

\begin{figure}
\begin{center}
\epsfig{file=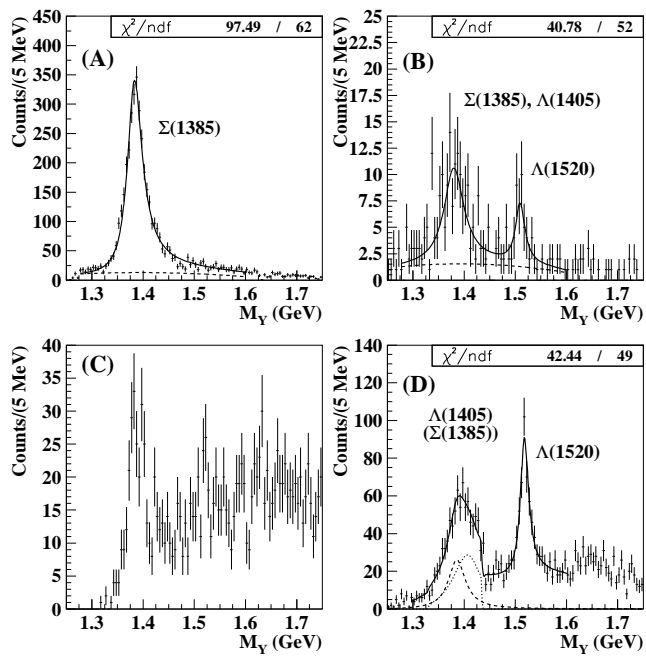,width=\linewidth}
\end{center}
\caption{Hyperon mass distributions for $\chi^2_{LOW}=
\chi^2_{HIGH}=3.841$ with  $p_\perp^2=0.015\ GeV^2$ cut.
The labels are explained in the text. The yield of $\Lambda\pi^0$ and  $\Lambda\gamma$ 
events in A and B were extracted by fitting the data with a 
relativistic Breit-Wigner (solid line) and a polynomial background (dashed line).
In D the dashed histogram shows the contribution 
due to the $\Sigma(1385)$ alone.  The dotted histogram 
is the $\Lambda(1405)$ contribution alone using the M-matrix parameterization
for the shape.}
\label{ystar_perp_cuts}
\end{figure}

This misidentification should happen for ground-state $\Sigma^0(1193)$ production as 
well. 
A subset of the data
filtered on the hyperon mass region between 1.0 and 1.25 GeV, was used to isolate 
$\Sigma^0(1193)$ events.  The $\sigma$ from a Gaussian fit to 
the $\Sigma^0$ peak is about 6.6 MeV, corresponding to a full width at half 
maximum of $\Gamma=2.354\sigma=15.6$ MeV.  This is a measure of the hyperon
mass resolution.
Apart from the hyperon mass range,  
the same set of cuts was used to analyze these data as for the excited-state
sample.  Fig. \ref{pperp}C shows the distribution in $p_\perp^2$
for this data set.
Figures \ref{pperp}B and \ref{pperp}D compare the effect of two choices for the 
$p_\perp^2$ cut on the hyperon mass distribution for the case where the $\gamma$ 
channel is favored.  The histograms  show the 
distributions in hyperon mass for those events that were cut out.  Histograms 
\ref{pperp}B and \ref{pperp}D both look like exponentially falling 
distributions.
Fig. \ref{ystar_perp_cuts} shows the 
corresponding hyperon mass spectra after applying the $p_\perp^2=0.015\ GeV^2$ cut. The histogram now
shows the expected structure for the $\Sigma(1385) \rightarrow \Lambda \gamma$ and $\Lambda(1520) \rightarrow
\Lambda \gamma$ reactions. Comparison with \ref{ystar_cuts}A shows that this cut
also reduces the number of $\Lambda \pi^0$ events seen. The Monte Carlo simulation
\ref{MC} is used to correct for this reduction.
The $p_\perp^2=0.015\ GeV^2$ cut will be used for the rest of the analysis.

\section{Acceptance}
\label{MC}

A detailed Monte Carlo simulation of the CLAS detector was performed using 
GEANT 3.21 for each of the three electron beam energies.
Table \ref{acc2} lists the set of reactions for which we generated events.
The experimental photon energy distribution was used   
to determine the energies of the incident photons in the 
simulation.
Relativistic Breit-Wigner shapes were used for the $\Sigma(1385)$, 
$\Lambda(1520)$  and 
$K^*$ mass distributions.  
For the $\Lambda(1520)$ the exponential slope for the t-dependence was 2.0 
$GeV^{-2}$.  The angular distribution for the radiative decay of the 
$\Lambda(1520)$ in its rest frame was taken to be proportional to $5-3\cos^2\theta$ 
according to the result obtained by Mast, et al.\cite{mast}. The same 
distribution was used for the $\Sigma\pi$ channels and 
for the $\Sigma(1385)$ decays.  
The model of Nacher, et al. \cite{Oset} with a flat angular distribution
was used for the $\Lambda(1405)$ decay channels. 

\begin{figure}
\begin{center}
\epsfig{file=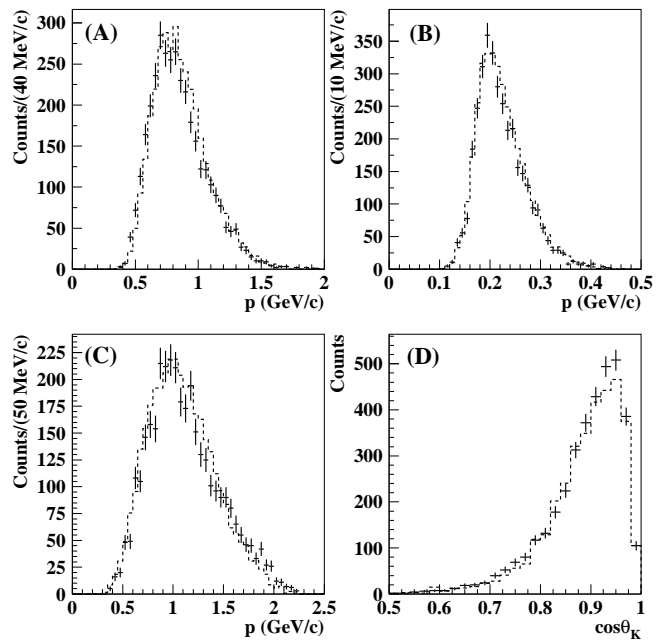,width=\linewidth}
\end{center}
\caption{Momentum and angular distributions for MC (dashed histograms) and 
data (points with error bars) for the 1.34--1.43 GeV hyperon mass region.}
\label{compmc}
\end{figure}

The incident photon energy dependence and 
t-dependence were adjusted to fit the data for the $\Sigma(1385)$ reactions 
independently for each of the electron beam energies. 
The data and MC were cut on the $Y^*$ mass range of 
1.34--1.43 GeV and on the $\pi^0$ peak found in Fig. \ref{ystar_perp_cuts}A to isolate
the $\Sigma(1385)\rightarrow \Lambda\pi^0$ channel. We plotted the ratio of the
data/MC versus photon energy $E_{beam}$. The resultant curve was fitted 
with a function of the form $A/E_{beam} + B/E_{beam}^2$.  We used this to 
modify the photon energy dependence of the $\Sigma(1385)$ production cross 
section in the MC.  The above procedure was then iterated. 
The exponential slope parameter was varied until the MC and 
data $t$ distributions matched reasonably well.
The exponential slope for the modified 
t-dependence was 1.0 $GeV^{-2}$. 
To check the quality of the simulation, we compared the momentum 
distributions 
for the Monte Carlo and the data for the kaon, proton, and pion tracks.
The simulated events were analyzed  
with the same cuts described above.
The results for the second iteration for the MC simulation are shown in 
Fig. \ref{compmc}.
The agreement between the MC and the data for the pion, proton, and kaon
momenta and the kaon lab angle is good.  

\begin{figure}
\begin{center}
\epsfig{file=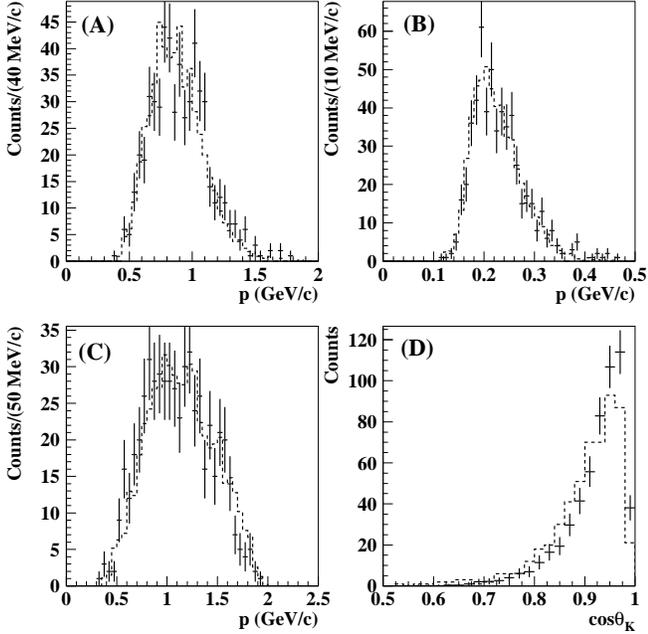,width=\linewidth}
\end{center}
\caption{Momentum and angular distributions for MC (dashed histograms) and 
data (points with error bars) for the 1.49--1.55 GeV hyperon mass region.}
\label{complammc}
\end{figure}

Fig. \ref{complammc} compares the data for the 1.49--1.55 GeV mass range
and the missing mass squared in 
the range 0.018--0.075 $GeV^2$ to the $\Lambda(1520)\rightarrow\Sigma^0\pi^0$ 
Monte Carlo results.  The MC results have been scaled by 0.185.  The agreement
between the MC and the measured  momenta distributions is very good and the 
kaon angular distributions agree reasonably well.

In order to check that the $p_\perp^2$ cut did not introduce a bias of the
Monte Carlo results with respect to the data, we studied the yield of 
$\Sigma(1385)\rightarrow\Lambda\pi^0$ events in the data and the corresponding
Monte Carlo.  For the data we used the standard $\chi^2$ cuts 
and performed the same kind of fit to the hyperon mass distributions as 
described earlier.
The hyperon mass range was 1.34--1.43 GeV.  The 
results are tabulated in table \ref{yields}.
\begin{table}
\begin{center}
\begin{tabular}{cccc}
$p_\perp^2$ cut $(GeV^2)$ &  N(data) & N(MC) &  N(data)/N(MC) \\\hline
0.005   &    4021   &   11037  & 0.364$\pm$0.007 \\
0.010   &    3500   &   9860   & 0.355$\pm$0.007 \\
0.015   &    2878   &   8148   & 0.353$\pm$0.008 \\
0.020   &    2191   &   6191   & 0.354$\pm$0.009 
\end{tabular}
\end{center}
\caption{Comparison of yields between the $\Sigma(1385)\rightarrow\Lambda\pi^0$
simulation and the data as a function of the $p_\perp^2$ cut. The errors are statistical only.}
\label{yields}
\end{table}
The data and the MC yields agree as a function of the $p_\perp^2$ cut. 

Table \ref{acc2} lists the acceptances for the case 
where $\chi^2_{HIGH}=\chi^2_{LOW}=3.841$.
\begin{table}
\begin{center}
\begin{tabular}{lcccc}
Reaction & $A_\pi$ & $A_\gamma$ & $A_{\gamma\pi}$ \\ \hline
$\Lambda(1405)\rightarrow\Sigma^0\pi^0$
       & 0.083$\pm$0.004 & 0.0007$\pm$0.0004 & 0.658$\pm$0.012 \\
$\Lambda(1405)\rightarrow\Sigma^+\pi^-$
       & 0.088$\pm$0.005 & 0.0038$\pm$0.0009 & 0.013$\pm$0.002   \\   
$\Lambda(1405)\rightarrow\Lambda\gamma$
       & 0.008 $\pm$0.003 & 0.946$\pm$0.028 & 0.098$\pm$0.009 \\
$\Lambda(1405)\rightarrow\Sigma^0\gamma$
       & 0.585$\pm$0.019 & 0.380$\pm$0.015 & 0.837$\pm$0.023 \\\hline
$\Sigma(1385)\rightarrow\Lambda\pi$    
       & 0.905$\pm$0.010 & 0.011$\pm$0.001 & 0.086$\pm$0.003 \\
$\Sigma(1385)\rightarrow\Sigma^+\pi^-$ 
       & 0.050$\pm$0.002 & 0.0018$\pm$0.0005& 0.00564$\pm$0.0008 \\
$\Sigma(1385)\rightarrow\Lambda\gamma$ 
       & 0.012$\pm$0.002 &  1.309$\pm$0.022 &  0.105$\pm$0.006 \\
$\Sigma(1385)\rightarrow\Sigma^0\gamma$
       &  0.548$\pm$0.016 & 0.24$\pm$0.01  & 0.99$\pm$0.02 \\\hline
$\Lambda(1520)\rightarrow\Lambda\gamma$ 
       &                  &   1.388$\pm$0.027     & 0.0010$\pm$0.0007 \\
$\Lambda(1520)\rightarrow\Sigma^0\gamma$
       &                  &   0.087$\pm$0.006     & 0.586$\pm$0.016 \\
$\Lambda(1520)\rightarrow\Lambda\pi^0\pi^0$
       &                 &      0        &  0.0099$\pm$0.0016 \\
$\Lambda(1520)\rightarrow\Sigma^0\pi^0$
      &                & 0.0006$\pm$0.0004  & 0.681$\pm$0.014 \\
\end{tabular}  
\end{center}
\caption{Acceptances (in units of $10^{-3}$) for the channels used in the 
calculation of the branching ratios. 
Here $\chi^2_{HIGH}=\chi^2_{LOW}=3.841$.
The uncertainties are 
statistical only.}
\label{acc2}
\end{table}
In the table $A_\pi$ and $A_\gamma$ refer to the fraction of surviving events 
relative to the number of thrown events that satisfy the $\Lambda\pi^0$ and
$\Lambda\gamma$ hypotheses, respectively, and $A_{\gamma\pi}$ refers to those 
events that do not satisfy either hypothesis.

\section{$\Sigma^0(1385)$ Analysis}
\label{sigma}

To obtain the yields we 
fitted the hyperon mass distributions between 1.25 GeV and 1.75 GeV. 
The yield of $\Lambda\pi^0$ events is extracted by fitting the data in 
 Fig. \ref{ystar_perp_cuts}A with a polynomial background and a
relativistic Breit-Wigner of the form \cite{jackson}
\begin{eqnarray}
f(M)\propto \frac{2 M M_0 \Gamma(q) }{(M^2-M_0^2)^2+M_0^2\Gamma^2(q)},\\
\Gamma(q)=\Gamma_0 \left( \frac{q}{q_0}\right)^{2l+1}\frac{M_0}{M} \left(
\frac{X_0^2+q_0^2}{X_0^2+q^2}\right)^l
,\\
q={\sqrt{(M^2-M_\Lambda^2-M_\pi^2)^2-4M_\Lambda^2 M_\pi^2} \over 2M}, \\
q_0={\sqrt{(M_0^2-M_\Lambda^2-M_\pi^2)^2-4M_\Lambda^2 M_\pi^2} \over 2M_0},
\end{eqnarray}
where $M_0$ is the peak position of the resonance, $X_0$=0.35 GeV and  $\Gamma_0$ is the width.  For the $\Sigma^*\rightarrow\Lambda\pi^0$ transition, $l=1$. 
 We 
tried both first order and second order polynomial background 
parameterizations. The systematic uncertainty in the yield extraction due to 
the choice of background function was about $\pm$1\%.  The 
mass and width of the $\Sigma(1385)$ were found to be 1.3860 GeV and 0.03988 GeV.
For the $\Lambda\gamma$ channel (Fig. \ref{ystar_perp_cuts}B),
we used two relativistic Breit-Wigners (one for the $\Sigma(1385)$ and 
one for the $\Lambda(1520)$) plus a polynomial background.  
The masses and widths were fixed to be those found from the fits to Figures \ref{ystar_perp_cuts}A 
and \ref{ystar_perp_cuts}D.

From Fig.\ref{ystar}A it is clear that we were not able 
to resolve the $\Lambda(1405)$ and the $\Sigma^0(1385)$, therefore 
in order to find the number of $\Lambda(1405)$'s ($n_\Lambda$) we look at the events 
for which neither the $\gamma$ nor the $\pi^0$ hypothesis is satisfied (Fig. \ref{ystar_cuts}D). This
isolated predominantly $\Lambda(1405)\rightarrow\Sigma^0\pi^0$ events, since the  
$\Sigma(1385) \rightarrow \Sigma^0\pi^0$ decay is forbidden by isospin.
We parameterized the $\Lambda(1405)$ 
line shape using the M-matrix formalism for S-wave $\Sigma^0\pi^0$ scattering
below the $\overline{K}N$ threshold.
The M-matrix is related to the S-wave transition matrix $T$ according to

\begin{equation}
T=Q^\frac{1}{2}(M-i Q)^{-1}Q^\frac{1}{2},
\end{equation}
where $Q$ is a diagonal matrix containing the relative $\Sigma^0\pi^0$ momentum
$q$ and $\overline{K}N$ momentum $k$ \cite{Thomas}.  Note that below the $\overline{K}N$ 
threshold, the latter is purely imaginary.  The matrix $M$ is expanded 
relative to the $\overline{K}N$ threshold $E_t=M_{\overline{K}}+M_{N}$
according to
\begin{eqnarray}
M(E)=M(E_t)+\frac{1}{2}R(Q^2(E)-Q^2(E_t)) \nonumber \\
 = \left[\begin{array}{cc} M_{11} & M_{12} \\ M_{12} & M_{22}
\end{array}\right], \\
Q=\left[ \begin{array}{cc} k & 0 \\ 0 & q \end{array}\right], \\
R=\left[ \begin{array}{cc} R_{\overline{K}N} & 0 \\ 0 & R_{\Sigma\pi} 
\end{array}\right].
\end{eqnarray}  
The amplitude for elastic scattering in the $\Sigma^0\pi^0$ channel is given by
\begin{equation}
T_{22}={q(M_{11}+|k|) \over 
    (M_{11}+|k|)(M_{22}-iq)-M_{12}^2}.
\end{equation}

Below $E_t$, the $\Sigma\pi$ mass spectrum is proportional to $|T_{22}|^2/q$.
Fig. \ref{ystar_perp_cuts}D shows the M-matrix parameterization fit to hyperon 
mass spectrum.  A relativistic Breit-Wigner form is included
to account for the leakage of the $\Sigma(1385)\rightarrow\Lambda\pi^0$ channel
into the high missing mass squared region.  A second relativistic Breit-Wigner 
is used for the $\Lambda(1520)$ contribution. The mass and width of the $\Lambda(1520)$
were found to be 1.520 GeV and 0.022 GeV.  We used a second-order 
polynomial for the remaining background beneath the peaks.
The matrix elements at threshold and the effective ranges were determined from
the fit to be $M_{11}(E_t)=1.314$, $M_{12}(E_t)=-1.063$, $M_{22}(E_t)=0.686$,
$R_{\overline{K}N}=9.543$, and $R_{\Sigma\pi}=-28.89$.
We find 328$\pm$36 $\Lambda(1405)$ counts and 245$\pm$37 
$\Sigma(1385)$ counts in the hyperon mass region 1.34--1.43 GeV. The reduced 
$\chi^2$ for the fit was 0.866.

\begin{table}
\begin{center}
\begin{tabular}{lc}
Reaction & Yield \\ \hline 
Estimated $\Sigma^0\pi^0$ counts & 373.8$\pm$34.0 \\  
\hline \hline
Raw $\pi^0$ counts & 2878.3$\pm$77.4  \\
$\Lambda(1405) \rightarrow \Sigma^0\gamma$ & 0.45$\pm$0.17 \\ 
$\Lambda(1405) \rightarrow \Sigma\pi$ & 95.7$\pm$9.5 \\ 
$\Sigma(1385) \rightarrow \Sigma\pi$ & 10.4$\pm$1.0 \\
$\Sigma(1385) \rightarrow \Lambda\gamma$ & 0.87$\pm$0.21 \\ 
\hline
Corrected $\pi^0$ counts & 2770.9$\pm$78.0 \\ 
\hline \hline
Raw $\gamma$ counts & 100.2$\pm$15.4 \\
$\Sigma(1385) \rightarrow \Lambda\pi^0$ & 35.0$\pm$1.0 \\ 
$\Sigma(1385) \rightarrow \Sigma^+\pi^-$ & 0.38$\pm$0.3 \\ 
$\Lambda(1405) \rightarrow \Lambda^0\gamma$ & 0.85$\pm$0.27 \\ 
$\Lambda(1405) \rightarrow \Sigma^0\gamma$ & 0.29$\pm$0.11 \\ 
$\Lambda(1405) \rightarrow \Sigma\pi$ & 2.47$\pm$0.25  \\ 
\hline
Corrected $\gamma$ counts & 61.2$\pm$15.4 
\\ \hline
\end{tabular}
\end{center}
\caption{Breakdown of statistics for the $\Lambda\gamma$ and $\Lambda\pi^0$ channels.  
The errors are statistical only.}
\label{statistics}
\end{table}

Although the $\pi^0$ leakage into the $\gamma$ channel is the dominant 
correction 
to the branching ratio, the final result still needs corrections for 
$\Sigma^+\pi^-$ contamination and the contribution to the numerator from the
reaction $\Lambda(1405)\rightarrow \Lambda\gamma$.  Based on the measured 27$\pm$8 keV 
radiative width\cite{Burkhardt},
we assume that the leakage of the $\Sigma\gamma$ 
channel into the $\gamma$ region is small relative to the 
$\Lambda\gamma$ signal and that the leakage into the $\pi^0$ region is 
small compared to the $\Lambda\pi^0$ signal.
The formula for the acceptance corrected branching ratio is
\begin{eqnarray}
R&=&\frac{1}{\Delta n_\pi A^\Sigma_\gamma
(\Lambda\gamma)
- \Delta n_\gamma A^\Sigma_\pi(\Lambda\gamma)} \nonumber \\
& & \times
\left[\Delta n_\gamma
 \left(A^\Sigma_\pi(\Lambda\pi)+\frac{R^{\Sigma\pi}_{\Lambda\pi}}{2}
A^\Sigma_\pi(\Sigma\pi)\right)\right. \nonumber \\
& & \left.-\Delta n_\pi
 \left(A^\Sigma_\gamma(\Lambda\pi)+\frac{R^{\Sigma\pi}_{\Lambda\pi}}{2}
A^\Sigma_\gamma(\Sigma\pi)\right)\right], \\
\Delta n_\pi&=&n_\pi-N_\pi(\Lambda^*\rightarrow \Sigma^+\pi^-)
                   -N_\pi(\Lambda^*\rightarrow \Sigma^0\pi^0) \nonumber \\
       & &         -N_\pi(\Lambda^*\rightarrow \Sigma^0\gamma)  
                   -N_\pi(\Lambda^*\rightarrow \Lambda \gamma), \\
\Delta n_\gamma&=&n_\gamma-N_\gamma(\Lambda^*\rightarrow \Sigma^+\pi^-)
                   -N_\gamma(\Lambda^*\rightarrow \Sigma^0\pi^0) \nonumber\\
        & &        -N_\gamma(\Lambda^*\rightarrow \Sigma^0\gamma)
		   -N_\gamma(\Lambda^*\rightarrow \Lambda \gamma),
\end{eqnarray}
where $n_\gamma$ ($n_\pi$) is the measured number of photon (pion) candidates 
and the remaining $N_{\gamma,\pi}$ terms are corrections due to leakage from 
the $\Lambda(1405)$.
The acceptance for the individual pion (photon) channels are denoted as 
$A^\Sigma_{\pi}(\Sigma^+\pi^-)$, ($A^\Sigma_{\gamma}(\Sigma^+\pi^-)$) and so on.
 For example, 
$A^\Sigma_\gamma(\Lambda\pi)$ denotes the relative leakage of 
the $\Lambda\pi$ channel into the $\Lambda\gamma$ channel. Table \ref{acc2} 
lists 
the values of these ``acceptances''. 

 The corrections depend on an estimate of 
the number $n_\Lambda$ of $\Lambda(1405)$'s in the data set.  They are
{\setlength\arraycolsep{2pt}
\begin{eqnarray}
N_{\gamma}(\Lambda^*\rightarrow\Lambda\gamma)&=&
 \frac{ A^\Lambda_{\gamma}(\Lambda\gamma) R(\Lambda^*\rightarrow\Lambda\gamma)
 n_\Lambda}{ A^\Lambda_{\gamma\pi}(\Sigma^0\pi^0)
 +A^\Lambda_{\gamma\pi} (\Sigma^+\pi^-)},\\
N_{\gamma}(\Lambda^*\rightarrow\Sigma^0\gamma)&=&
 \frac{ A^\Lambda_{\gamma}(\Sigma^0\gamma) R(\Lambda^*\rightarrow\Sigma^0\gamma)
 n_\Lambda}{A^\Lambda_{\gamma\pi}(\Sigma^0\pi^0)
 +A^\Lambda_{\gamma\pi} (\Sigma^+\pi^-)},\\
N_{\gamma}(\Lambda^*\rightarrow\Sigma^0\pi^0)&=&
\frac{ A^\Lambda_{\gamma}(\Sigma^0\pi^0) n_\Lambda
 }{ A^\Lambda_{\gamma\pi}(\Sigma^0\pi^0)
 +A^\Lambda_{\gamma\pi} (\Sigma^+\pi^-)},\\
N_{\gamma}(\Lambda^*\rightarrow\Sigma^+\pi^-)&=&\frac
{ A^\Lambda_{\gamma}(\Sigma^+\pi^-) n_\Lambda
 }{ A^\Lambda_{\gamma\pi}(\Sigma^0\pi^0)
 +A^\Lambda_{\gamma\pi} (\Sigma^+\pi^-)},
\end{eqnarray}}
and similarly for the pion channel.  Here isospin symmetry is assumed such 
that $R(\Sigma^0\pi^0)=R(\Sigma^+\pi^-)=R(\Sigma^-\pi^+)\approx 1/3$ for the 
$\Lambda(1405)$ decay channels.  The subscript ``$\gamma\pi$'' refers to those
events for which both a pion and a photon are missing or those events leaking 
into the ``$\gamma\pi$'' region due to the tail of the $\pi^0$ peak (this is 
why the $\Sigma^+\pi^-$ 
contamination must be included in the denominator, although the leakage for 
this channel is small).  
Table \ref{statistics} lists the yields for the various channels 
of the $\Sigma(1385)$ decays. 
The hyperon mass range was 1.34--1.43 GeV.
The reaction $\gamma p \rightarrow \Lambda K^{*+}$ causes a smooth background underneath the 
$\Sigma(1385)$ peak in Figures \ref{ystar_perp_cuts}A and \ref{ystar_perp_cuts}B that is well 
parameterized by the second order polynomial 
fit. Hence it has not been explicitly included Table \ref{statistics}.
The largest background in the 
$\gamma$ channel is due to leakage of the $\pi^0$ tail into the $\gamma$ 
missing mass squared region.

After subtracting the background contributions enumerated in table \ref{statistics}  
there were $61.2\pm15.4$ counts consistent with 
$\Sigma^0(1385)\rightarrow\Lambda\gamma$
and $2770.9\pm78.0$ counts consistent with $\Sigma^0(1385)\rightarrow\Lambda\pi^0$.
After correcting for the relative acceptance of the two channels,
we obtained a branching ratio,  
$R^{\Lambda\gamma}_{\Lambda\pi}$, of
\begin{equation}
\frac{\Gamma(\Sigma^0(1385)\rightarrow\Lambda\gamma)}
     {\Gamma(\Sigma^0(1385)\rightarrow\Lambda\pi^0)}
=1.53\pm0.39(\mathrm{stat})\%.
\end{equation} 

\begin{figure}
\begin{center}
\epsfig{file=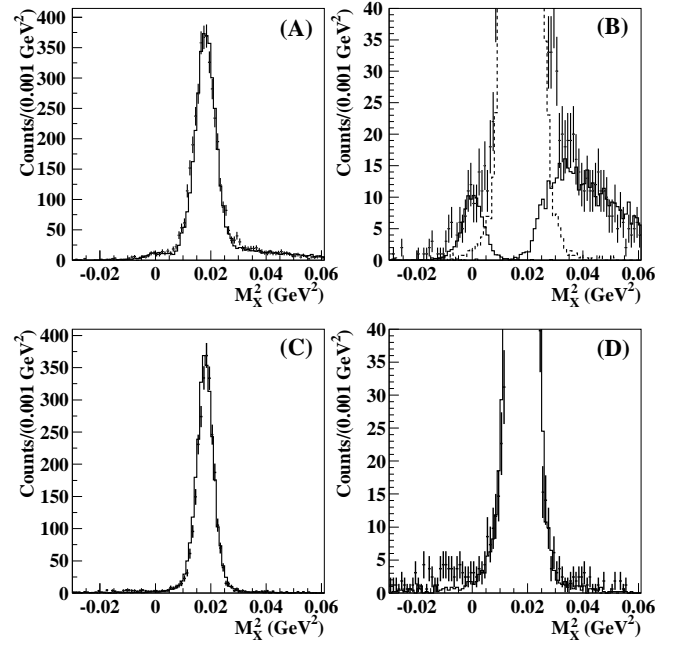,width=\linewidth}
\end{center}
\caption{Comparison between data and Monte Carlo results for the reactions
 and $Y^*\rightarrow\Lambda X$ (top histograms) $Y^* \rightarrow \Sigma^+\pi^-$
 (bottom histograms) after kinematic fitting has been performed.  The points 
with error bars are the data and the curves are the MC results. Histrograms B and D have the vertical scales expanded by a factor of ten.
In B the solid curve on the left is the $\Sigma^*\rightarrow\Lambda\gamma$ 
simulation, the central dashed curve is the $\Sigma^*\rightarrow\Lambda\pi^0$ 
simulation, the isolid curve on the right is the $\Lambda(1405)\rightarrow\Sigma^0\pi^0$
simulation. In A the curve is the sum of the three.
In C and D the $\Sigma^+\pi^-$ data and
the $\Sigma(1385)\rightarrow\Lambda\pi^0$ Monte Carlo distribution have been 
scaled to agree with the peak height of the $\pi^0$ in the 
$Y^*\rightarrow\Lambda X$ distribution from the data set. }
\label{sigp2}
\end{figure}

The branching ratio result for the $\Sigma(1385)$ depends on how well we
understand the tail of the $\pi^0$ peak near the $\gamma$ peak. 
Fig. \ref{sigp2}C and \ref{sigp2}D shows the comparison between the data and
the Monte Carlo for the reaction $Y^*\rightarrow\Lambda X$ for the 1.34--1.43 GeV 
hyperon mass region.  The excess of counts above the $\pi^0$ peak correspond to 
$Y\to\Sigma^0\pi^0$, where $\Sigma^0\to\Lambda\gamma$. Although the 
$Y\to\Lambda \gamma$ decay is not
completely separated, a clear enhancement near zero missing mass squared 
can be seen above the $\pi^0$ tail 
clearly indicating the presence of radiative events. 
The Monte Carlo predicts that the leakage accounts
for about 30\% of the raw photon yield in the $|M_X^2|<0.01$ GeV$^2$ region. 
 In order to assess
the quality of the Monte Carlo in the tail, we looked at 
$\Lambda(1405)\rightarrow
\Sigma^+\pi^-$ events for which the $\Sigma^+$ subsequently decayed to 
$p\pi^0$.  We chose this channel because there are no 
channels that can distort the spectrum above the $\pi^0$ peak, the
$\Sigma^+$ radiative channel is rare ($BR=1.25\times 10^{-3}$), and has similar kinematics 
to the $\Lambda \pi^0$ decay.
 We required the $p\pi^-$ invariant mass to be greater than
1.13 GeV (to eliminate the $\Lambda(1116)$ from the sample).  To identify the 
$\Sigma^+(1189)$ we require
the $pX$ invariant mass (or, equivalently, the missing mass recoiling off the  
$K^+\pi^-$ system) to 
be in the range 1.17-1.206 GeV. We performed 
kinematic fits on these events with vertex
and four-momentum conservation constraints. Explicitly, the constraint 
equations are 
\begin{equation}
\vec f = \left[ \begin{array}{c}
	E_{beam} + m_p - E_K - E_\pi-E_{\Sigma^+} \\
	\vec p _{beam} -  \vec p _\pi - \vec p _p - \vec p_{\Sigma^+} \\
	(y-y_\pi) p^z_\pi-(z-z_\pi)p^y_\pi\\
        (x-x_\pi) p^z_\pi-(z-z_\pi)p^x_\pi\\
        (y-y_p) p^z_K-(z-z_p)p^y_K\\
        (x-x_p) p^z_K-(z-z_p)p^x_K\\
	\end{array} \right]=\vec 0. 
\end{equation} 
The missing mass squared distribution for the reaction chain
$Y^*\rightarrow \Sigma^+ \pi^-$, $\Sigma^+ \rightarrow p (X)$ is shown in 
Fig. \ref{sigp2}C and \ref{sigp2}D for hyperon masses in the 1.38--1.45 GeV 
mass 
region.  We used the four-vector for the $\Sigma^+$ obtained from the fit, with 
less than 0.5\% probability of exceeding $\chi^2$.  The Monte Carlo result (dashed histograms)
for the $\Lambda(1405)\rightarrow\Sigma^+\pi^-$ reaction agrees
very well with the data down to about zero missing mass squared.
The discrepancy between the MC and the data in the $-$0.01 -- +0.01 $GeV^2$ region is about $\sim19\%$.  
Scaling the leakage of the $\Lambda\pi^0$ channel into the $\gamma$ region by 
a factor of 1.19 reduces the branching ratio from 1.53\% to 1.36\% for a 
relative change of about $-$11\%.   More importantly, comparing \ref{sigp2}B with 
\ref{sigp2}D shows a clear enhancement at zero missing mass present for the 
latter case not in evidence for the former case.
The negative systematic error will be increased by 11\% in 
quadrature. 

\section{$\Lambda(1520)$ analysis} 

\begin{figure}
\begin{center}
\epsfig{file=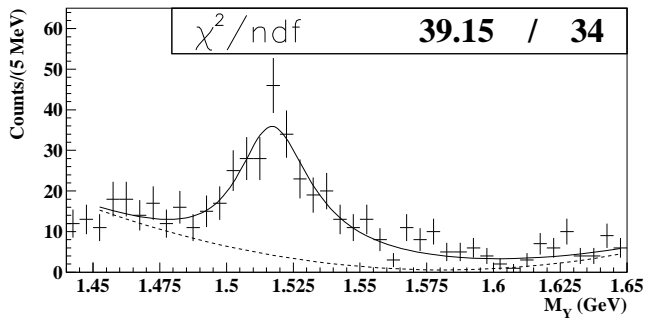,width=\linewidth}
\end{center}
\caption{Sample fit of the $\Lambda X$ mass distribution for missing mass 
squared in the 0.018-0.075 $GeV^2$ range.}
\label{lam1520fit}
\end{figure}

\begin{figure}
\begin{center}
\epsfig{file=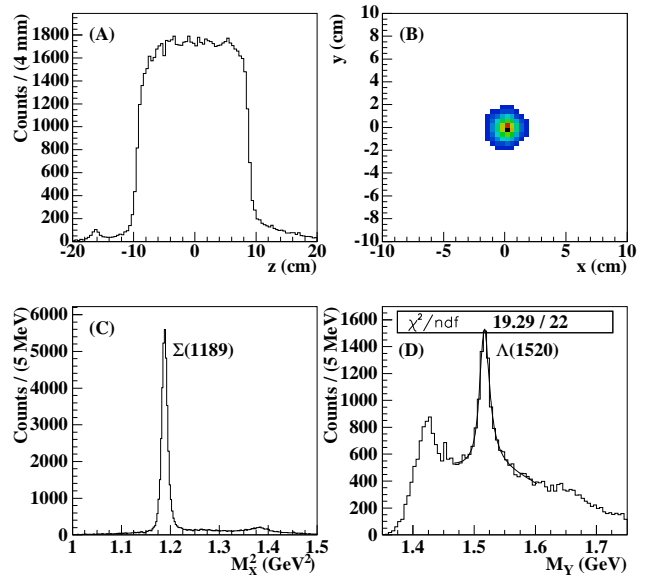,width=\linewidth}
\end{center}
\caption{Isolation of $Y^*\rightarrow\Sigma^+\pi^-$ events.  A) and B) show
the $K^+\pi^-$ vertex distributions.  C) is the missing mass for the 
reaction $\gamma p \rightarrow K^+\pi^- X$.  D) is the hyperon mass 
distribution for events satisfying the $\Sigma^+$ identification cut 
(see text).}
\label{lam1520sigpi}
\end{figure}

For the $\Lambda(1520)$ analysis 
we calculated the radiative branching ratio relative to the $\Sigma^0\pi^0$ 
and the $\Sigma^+\pi^-$ channels.
 The hyperon mass cut used to identify the 
$\Lambda(1520)$ was 1.49-1.55 GeV.  From the fit to the histogram shown in
Fig. \ref{ystar_perp_cuts}, we obtained $n_\gamma=32.5\pm8.2$.
 To identify the $\Sigma^0\pi^0$ channel
we used events for which neither the $\gamma$ 
nor the $\pi^0$ hypothesis is satisfied. 
The ground-state $\Lambda$ is a decay product in the
$\Sigma^0\pi^0$ (14\%) and $\Lambda\pi\pi$ (10\%) channels.  In order to simplify the
calculation for the branching ratio we require the missing mass squared 
to be in the range between $m_{\pi^0}^2$ and 0.075 $GeV^2$
($\approx4m_{\pi^0}^2$, the two-pion threshold).  This isolates the 
$\Sigma^0\pi^0$ channel.  The hyperon mass distribution in the $\Lambda(1520)$
region with this additional cut applied is shown in Fig. \ref{lam1520fit}.
The fit is a D-wave ($l=2$) relativistic Breit-Wigner plus a polynomial 
background.  We tried both first-order and second-order polynomials; 
the results for the yield differed by $\pm1.6\%$. 
\begin{table}
\begin{center}
\begin{tabular}{lc}
 Reaction & Yield \\ \hline 
$\Lambda(1520) \rightarrow \Sigma^+\pi^-$  & 5290$\pm$124  
\\ \hline
$\Lambda(1520) \rightarrow \Sigma^0\pi^0$  & 202.8$\pm$16.7 \\  
$\Lambda(1520) \rightarrow \Lambda\gamma$ & 0.05$\pm$0.01
\\ \hline
 Corrected $\pi^0$ counts & 202.8$\pm$16.7 
\\ \hline \hline
 Raw $\gamma$ counts & 32.5$\pm$8.2 \\
$\Lambda(1520) \rightarrow \Sigma^0\pi^0$ & 0.09$\pm$0.01  
\\ \hline
 Corrected $\gamma$ counts & 32.4$\pm$8.2 
\\ \hline
\end{tabular}
\end{center}
\caption{Breakdown of statistics for the $\Lambda(1520)$ analysis.  
The errors are statistical only.}
\label{lam1520_stat}
\end{table}
The leakage of one channel into the other is neglible and applying the correction
does not change the result.
Due to the low acceptance for events containing $\Lambda$'s, the raw number of 
$\Sigma^0 \pi^0$ counts is only a factor of 6 larger than the radiative 
signal and the technique relies on isolating a channel for which two particles 
($\gamma$ and $\pi^0$) are not detected.  We also looked at 
$\Lambda(1520)\rightarrow\Sigma^+\pi^-$ events for which the acceptance is 
higher.  The same particle identification and vertex cuts used for the
previous analysis were applied with some modifications.  We required that 
the $p\pi^-$ invariant mass be greater than 1.13 GeV to cut $\Lambda(1116)$
contamination.  The primary vertex was determined using the $K^+$ and $\pi^-$
tracks.  The z-position and x- and y-positions for these vertexes are shown
in Fig. \ref{lam1520sigpi}A and \ref{lam1520sigpi}B, respectively.
A prominent $\Sigma^+(1189)$ peak shows up in the missing mass recoiling 
against the $K^+$ and the $\pi^-$ (Fig. \ref{lam1520sigpi}C).  The hyperon 
mass spectrum for those events in the range 1.165--1.215 GeV about the 
$\Sigma^+$ peak are shown in Fig. \ref{lam1520sigpi}D. The curve is a fit to 
the $\Lambda(1520)$ region using a D-wave relativistic Breit-Wigner with a 
second-order polynomial background. In the region between 1.49 GeV and 1.55 GeV
we obtain $n_{\Lambda^*}=5290\pm124$ (the acceptance of CLAS is much
larger for this channel than the others due to the larger $\pi^-$ momentum). 
 The yields for these two reactions 
are listed in table \ref{lam1520_stat}.
As can be seen from the numbers in the table the leakage of each channel into the other is negligible.
None of the generated $\Lambda(1520)\rightarrow\Sigma^+\pi^-$ events 
satisfied the selection criteria.
There is no $\Lambda\pi^0$ leakage since this channel 
is forbidden by isospin.
 
We obtained a raw branching ratio of 
$n_\gamma/n_{\gamma\pi}=16.0\pm4.3\%$.  
Correcting for acceptance, the 
branching ratio is
\begin{equation}
{\Gamma(\Lambda\gamma) \over \Gamma(\Sigma^0\pi^0)}= 
\frac{A_{\gamma\pi}(\Sigma^0\pi^0)}
{A_\gamma(\Lambda\gamma)} \frac{n_\gamma}{n_{\gamma\pi}} = 7.9\pm2.1\%.
\end{equation}
The acceptances used in this calculation are listed in table \ref{acc2}.
To obtain the branching ratio $\Gamma(\Lambda\gamma)/\Gamma_{TOT}$ we scale 
this result by the branching fraction of 14\% for the $\Sigma^0\pi^0$ channel
(assuming isospin symmetry) to obtain 1.10$\pm$0.29\%.
The acceptance for the $\Lambda(1520)\rightarrow\Sigma^+\pi^-$ channel 
was $1.66\pm0.06$\%.  We obtain $1.01\pm0.26$\%  for the radiative branching 
ratio. The results for the two channels agree after acceptance corrections. 

If contamination due to the
$\Lambda(1520)\rightarrow \Sigma^0\gamma$ channel is present 
the branching ratio for the $\Lambda\gamma$ channel acquires a small
correction term:
{\setlength\arraycolsep{2pt}
\begin{eqnarray}
R(\Lambda\gamma)&=&\frac{A_{\gamma\pi}(\Sigma^0\pi^0)}{A_\gamma(\Lambda\gamma)} 
\frac{n_\gamma}{n_{\gamma\pi}} R(\Sigma^0\pi^0) \nonumber \\
 &   &     +R(\Sigma^0\gamma)
\left( { A_{\gamma\pi}(\Sigma^0\gamma)n_\gamma-A_\gamma(\Sigma^0\gamma)
n_{\gamma\pi} \over A_\gamma(\Lambda\gamma) n_{\gamma\pi}} \right),
\end{eqnarray}}
where $R(\Sigma^0\gamma)$ is the branching ratio to the $\Sigma^0\gamma$ 
channel and $R(\Sigma^0\pi^0)$ is the branching ratio to the $\Sigma^0\pi^0$ 
channel.  Using the largest theoretical estimate for the $\Sigma^0\gamma$ 
radiative width of 293 keV from Warns, \emph{et al.}\cite{warns}, we obtain
a correction of +0.01\%. Therefore this contamination can be neglected.

\section{Results}

To check the sensitivity to the confidence limits used, 
$R^{\Lambda\gamma}_{\Lambda\pi^0}$ was calculated with  1\%, 5\% and 10\% probability for accepting a channel and  
99\%, 95\% and 90\% probability for rejecting a channel. Table \ref{chi2_cuts} lists the corrected
branching ratios as a function of the $\chi^2_{HIGH,LOW}$ cuts. 
The third column in Table \ref{chi2_cuts} gives the $\Sigma(1385)$ results.
The results were very stable, varing from +0.15 (10\%, 90\%) to $-$0.17 (5\%, 99\%).  
These values were used as estimates of the systematic errors. 
The value for the branching ratio is 
$1.53\pm0.39(\mathrm{stat})^{+0.15}_{-0.17}(\mathrm{sys})$\%, where 
the second uncertainty reflects the variation in the branching ratio as 
a function of the choice of $\chi^2$ cuts.  

\begin{table} 
\begin{center}
\begin{tabular}{cccc}
$\chi^2_{LOW}$ & $\chi^2_{HIGH}$ & R(\%) $\Sigma(1385)$ & R(\%) $\Lambda(1520)$ \\ \hline
2.706 & 2.706 & 1.68$\pm$0.41 & 1.20$\pm$0.29 \\
3.841 & 3.841 & 1.53$\pm$0.39 & 1.10$\pm$0.29 \\
6.635 & 6.635 & 1.58$\pm$0.40 & 1.13$\pm$0.33 \\
2.706 & 6.635 & 1.38$\pm$0.38 & 1.25$\pm$0.33 \\
3.841 & 6.635 & 1.36$\pm$0.30 & 1.06$\pm$0.34
\end{tabular}
\end{center}
\caption{Dependence of the $\Sigma(1385)\rightarrow \Lambda \gamma$ and $\Lambda(1520)\rightarrow \Lambda \gamma$ branching 
ratio on the choice of $\chi^2_{HIGH,LOW}$ cuts.  
}
\label{chi2_cuts}
\end{table} 

We add the 11\% relative error (i.e. $-$0.17\% absolute) that could result 
from underestimating the tail of the $\pi^0$ response to the negative systematic error 
and quote a branching ratio of $1.53\pm0.39(\mathrm{stat})^{+0.15}_{-0.24}(\mathrm{sys})$\%.
The positive systematic error reflects the range of values 
we obtained for the various estimates for the branching ratio.
If we neglect the small (unmeasured) contribution due to the $\Sigma^0\gamma$
channel, the $\Sigma^0(1385)\rightarrow\Lambda\gamma$ partial width is given by
\begin{equation}
\Gamma(\Lambda\gamma)={ R^{\Lambda\gamma}_{\Lambda\pi} \Gamma_{TOT} \over
1+R^{\Lambda\gamma}_{\Lambda\pi}+R^{\Sigma\pi}_{\Lambda\pi}}
=479\pm120(\mathrm{stat})^{+81}_{-100}(\mathrm{sys})\mbox{ keV},
\end{equation}
using $\Gamma_{TOT}=36\pm5$ MeV and $R^{\Sigma\pi}_{\Lambda\pi}=0.135\pm0.011$,
the branching ratio of the $\Sigma\pi$ channels relative to the 
$\Lambda\pi^0$ channel\cite{pdg}.
The errors on $\Gamma_{TOT}$ and $R^{\Sigma\pi}_{\Lambda\pi}$ are included 
in the systematic error for $\Gamma(\Lambda\gamma)$.
If we use the largest theoretical estimate for the $\Sigma^0\gamma$ channel
relative to the $\Lambda\gamma$ channel of 0.153 from  R.~Bijker, F.~Iachello, and A.~Leviatan\cite{Bijker:2000gq}, the partial width is reduced to 478 keV, which 
is an insignificant change.  

For the $\Lambda(1520)$ decay, we obtained a branching ratio of 
$1.10\pm0.29(\mathrm{stat})^{+0.15}_{-0.04}(\mathrm{sys})$\% using the $\Sigma^0\pi^0$ channel 
 and $1.01\pm0.27$\% using the $\Sigma^+\pi^-$ channel.  
The weighted average gives a branching ratio of 
\begin{equation}
\frac{\Gamma(\Lambda(1520)\rightarrow\Lambda\gamma)}{\Gamma_{TOT}}=
1.07\pm0.29(\mathrm{stat})^{+0.15}_{-0.04}(\mathrm{sys})\%. 
\end{equation}
 Table \ref{chi2_cuts} lists the branching ratios for various combinations of
 kinematic fitting $\chi^2$ cuts.  There is no obvious 
dependence on the choice of cuts.  
 To determine the systematic error in the measurement
using the $\Sigma^0\pi^0$ channel to normalize, we used the range of branching 
ratio  values obtained for different choices of $\chi^2$ cuts. 
Using a full width of 15.6$\pm$1 MeV\cite{pdg}, we obtain a partial width of 
$167\pm43(\mathrm{stat})^{+26}_{-12}(\mathrm{sys})$ keV. The error on the full width is 
included in the systematic error for $\Gamma(\Lambda\gamma)$.
The $\Lambda(1520)$ result is compatible with the Mast\etal\
result\cite{mast} and the Antipov\etal\ result\cite{Antipov:2004qp} but 
disagrees with the Bertini\etal\ result\cite{bertini}.  
Together, our result and those of  
Mast\etal\ and Antipov\etal\ exclude the bag models listed in Table 
\ref{widths}.

The $\Sigma^0(1385)\rightarrow\Lambda\gamma$ channel has never been measured 
before.  The result is roughly 2--3 times larger than all of the existing model
 predictions except for HB$\chi$PT\cite{butler}.
Table \ref{widths} reveals that the model predictions for the 
$\Delta\rightarrow p \gamma$ transition are also about 50\% low. 
Sato and Lee\cite{sato}
showed that much of that discrepancy could be accounted for by the
inclusion of non-resonant meson-exchange effects. They find a width
of 530$\pm$45 keV, about 80\% of the experimental value.
Lu\etal\cite{lu}
reproduced the $\Delta\rightarrow p \gamma$ data using a chiral bag model 
calculation with a relatively small bag 
radius of 0.7 fm.  About 40\% of the transition was due to the pion 
cloud. These calculations suggest that mesonic effects could account for the 
discrepancy between the model predictions and our result for the $\Sigma^0(1385)$
radiative transition.

We would like to acknowledge the outstanding efforts of the staff of the 
Accelerator and the Physics Divisions at TJNAF that made this experiment 
possible.
This work was supported in part by the Istituto Nazionale di Fisica Nucleare, 
the  French Centre National de la Recherche Scientifique, 
the French Commissariat \`{a} l'Energie Atomique, the U.S. Department of
 Energy, the National Science Foundation, Emmy Noether grant from the Deutsche 
Forschungs Gemeinschaft and the Korean Science and Engineering Foundation.
The Southeastern Universities Research Association (SURA) operates the 
Thomas Jefferson National Accelerator Facility for the United States 
Department of Energy under contract DE-AC05-84ER40150.

\end{document}